\documentclass[fleqn,usenatbib]{mnras}

\usepackage{newtxtext,newtxmath}
\usepackage[T1]{fontenc}
\usepackage{ae,aecompl}

\usepackage{graphicx}	% Including figure files
\usepackage{subfig}
\usepackage{amsmath}	% Advanced maths commands
\usepackage{verbatim}
\usepackage{epstopdf}
\usepackage{multirow}
\usepackage{pifont}% http://ctan.org/pkg/pifont
\usepackage{array}

\usepackage[dvipsnames]{xcolor}

\newcommand{\g}{\rm{\gamma}}
\newcommand{\cmark}{{\color{blue}\LARGE\ding{51}}}%
\newcommand{\xmark}{{\color{red}\LARGE \ding{55}}}%

\newcommand{\integral}{\textit{INTEGRAL}}%\xspace}
\newcommand{\xmm}{\textit{XMM-Newton}}%\xspace}
\newcommand{\nustar}{\textit{NuSTAR}}%\xspace}
\title[Hadronic processes in Cygnus X--1]{A new lepto-hadronic model applied to the first simultaneous multiwavelength data set for Cygnus X--1}
\author[D. Kantzas et al.]{
D. Kantzas$^{1,2}$\thanks{E-mail: d.kantzas@uva.nl}, 
S. Markoff$^{1,2}$, 
T. Beuchert$^{3,1}$,
M. Lucchini$^{1}$, 
A. Chhotray$^{1}$, 
\newauthor
 ~C. Ceccobello$^4$,
A. J. Tetarenko$^5$,
J. C. A. Miller-Jones$^6$, 
 ~M. Bremer$^{7}$,
J. A. Garcia$^{8,9}$, 
\newauthor
~V. Grinberg$^{10}$,
P. Uttley$^{1}$
\&  J. Wilms$^9$ 
\\
$^{1}$Anton Pannekoek Institute for Astronomy (API), University of Amsterdam, Science Park 904, 1098 XH Amsterdam, the Netherlands\\
$^{2}$GRavitational AstroParticle Physics Amsterdam (GRAPPA), University of Amsterdam, Science Park 904, 1098 XH Amsterdam, the Netherlands\\
$^3$European Southern Observatory, Karl-Schwarzschild-Stra{\ss}e 2, 85748 Garching bei M{\"u}nchen, Germany\\
$^4$Department of Space, Earth and Environment, Chalmers University of Technology, Onsala Space Observatory, 439 92 Onsala, Sweden\\
$^5$East Asian Observatory, 660 N. A'{o}h\={o}k\={u} Place, University Park, Hilo, Hawaii 96720, USA\\
$^6$International Centre for Radio Astronomy Research - Curtin University, GPO Box U1987, Perth, WA 6845, Australia\\
$^7$Institut de Radio Astronomie Millim\'etrique (IRAM), 300 rue de la Piscine, 38406 Saint Martin d'H\`eres, France\\
$^8$Cahill Center for Astronomy and Astrophysics, Caltech, 1200 East California Boulevard, Pasadena, CA 91125, the USA\\
$^9$Dr. Karl Remeis-Observatory and Erlangen Centre for Astroparticle Physics, Sternwartstr. 7, 96049 Bamberg, Germany\\
$^{10}$Institute for Astronomy und Astrophysics, University of T{\"u}bingen, Sand 1, 72076 T\"{u}bingen, Germany\\
}

\date{Accepted XXX. Received YYY; in original form ZZZ}

\pubyear{2020}

\begin{document}
\label{firstpage}
\pagerange{\pageref{firstpage}--\pageref{lastpage}}
\maketitle

\begin{abstract}
%\vspace{8cm} %5mm vertical space
Cygnus X--1 is the first Galactic source confirmed to host an accreting black hole. It has been detected across the entire electromagnetic spectrum from radio to GeV $\g$-rays. The source's radio through mid-infrared radiation is thought to originate from the relativistic jets. The observed high degree of linear polarisation in the MeV X-rays suggests that the relativistic jets dominate in this regime as well, whereas a hot accretion flow dominates the soft X-ray band. The origin of the GeV non-thermal emission is still debated, with both leptonic and hadronic scenarios deemed to be viable. In this work, we present results from a new semi-analytical, multi-zone jet model applied to the broad-band spectral energy distribution of Cygnus X--1 for both leptonic and hadronic scenarios. We try to break this degeneracy by fitting the first-ever high-quality, simultaneous multiwavelength data set obtained from the CHOCBOX campaign (Cygnus X--1 Hard state Observations of a Complete Binary Orbit in X-rays). Our model parameterises dynamical properties, such as the jet velocity profile, the magnetic field, and the energy density. Moreover, the model combines these dynamical properties with a self-consistent radiative transfer calculation including secondary cascades, both of leptonic and hadronic origin. We conclude that sensitive TeV $\g$-ray telescopes like Cherenkov Telescope Array (CTA) will definitively answer the question of whether hadronic processes occur inside the relativistic jets of Cygnus X--1.   
\end{abstract}


%%%%%%%%%%%%%%%%%%%%%%%%%%%%%%%%%%%%%%%%%%%%%%%%%%
%%%%%%%%%%%%%%%%% BODY OF PAPER %%%%%%%%%%%%%%%%%%

\section{Introduction}
Throughout the Universe, a significant fraction of accreting black holes are known to launch relativistic and collimated jets. Fundamental properties, such as the extent and power of these jets, scale essentially with the mass of the central black hole. While super-massive black holes (SMBHs) with $\rm{M_{BH}}\sim 10^6$--$10^9\,{\rm{M_\odot}}$ located at the center of Active Galactic Nuclei (AGN) are able to power jets up to Mpc scales \citep[e.g.,][]{waggett1977ngc}, Galactic black holes ($\rm{M_{BH}\sim}$ tens of $\rm{M_{\odot}}$) hosted by X-ray binaries (XRBs) typically launch jets that remain collimated up to sub-pc scales  \citep[e.g.,][]{mirabel1994superluminal,hjellming1995episodic,mioduszewski2001one,gallo2005dark,fender2006transient,rushton2017resolved,russell2019maxij1535571}.

AGN jets carry enough power to accelerate particles up to ultra-high energies of $10^{19}$ eV and above \citep{aharonian2000tev}, which we detect as cosmic rays (CRs) on Earth. The exact acceleration mechanism is not known, but is likely related to diffusive shock acceleration \citep{axford1969acceleration,blandford1978particle,1990ApJ...360..702E,rieger2007fermi}, magnetic re-connection \citep{spruit2001large,giannios2010uhecrs,sironi2015relativistic}, or shearing and instabilities at boundary layers between different velocities \citep{rieger2004shear,liu2017particle}. 

The CR spectrum detected on Earth covers more than ten orders of magnitude in particle energy, from $10^{9}\,$ to $\sim 10^{21}\,$eV. Two well-known characteristic spectral features of that spectrum are the so-called 'knee' at $10^{15}\,$eV and the 'ankle' at $10^{18}\,$eV \citep[][respectively]{kulikov1959size,Bird1993ankle}. As shown by \cite{hillas1984origin}, the maximum energy of the accelerated particles at a given magnetic field is limited by the size of the source due to confinement arguments. Accordingly, CRs above the ankle are likely of extragalactic origin whereas CRs below the knee are of Galactic origin. AGN jets are considered the most likely source of extragalactic CRs \citep[e.g.,][and references therein]{hillas1984origin,gaisser2016cosmic,eichmann2018ultra}. Supernovae and supernova remnants have been considered the dominant source of Galactic CRs for decades although questioned quite recently due to lack of $\ge 100\,$TeV observations \citep{aharonian2018massive}. Hence, new candidate sources are needed. 

Large AGN jets and small-scale XRB jets are (self-)similar in many regards. For example, they display similar non-thermal emission processes, suggesting that both classes are capable of accelerating particles to high energies regardless of their physical scales \citep[e.g.,][]{markoff2001jet,bosch2006broadband,zdziarski2012mev}. Recent observations of hydrogen and helium emission lines from the jets of the accreting compact object SS~433 \citep{2004ASPRv..12....1F}, as well as the iron emission lines from the stellar-mass black hole candidate 4U~1630-47 \citep{trigo2013baryons}, provide indirect evidence of hadronic content their jets. It is still not clear whether XRB jets can efficiently accelerate hadrons to high energy, but if so, they could also be potential Galactic CRs sources \citep[see e.g.,][]{heinz2002CRmicroquasars,fender2005CRXRBs,cooper2020xrbcrs}.

The most striking evidence for particle acceleration inside Galactic jets comes from the non-thermal GeV radiation detected by the XRBs Cygnus X--1 (Cyg~X--1) and Cygnus X-3 \citep{malyshev2013high,bodaghee2013gamma,zanin2016detection,Tavani_2009CygnusX3}. The jet-origin of the  GeV emission is further favored by the orbital modulation predicted, e.g., by \citet{bottcher2005photon}. \citet{zdziarski2017high} in fact detected an MeV--GeV modulation that likely originates from synchrotron self-Compton upscattering by particles accelerated in the compact black-hole-jet system of Cyg~X--1 orbiting its companion star.

The exact nature of the non-thermal radiation is still unclear, with both leptonic and hadronic processes deemed to be viable. In the former case, a leptonic population is responsible for the overall electromagnetic spectrum from radio to $\g$-rays \citep[e.g.,][]{bosch2006broadband}. In the latter case, the hadronic population reaches relativistic speeds as well and contributes equally, or even dominates, in the high energy regime of the spectrum. According to the Hillas criterion, particles can attain high-enough energy only if a strong magnetic field confines them in the acceleration region and provides enough power for particle acceleration. The power carried by accelerated protons has been claimed to exceed the Eddington luminosity in several cases making the hadronic model controversial \citep{zdziarski2015hadronic}. The hadronic channel, however, is the only possible way to explain the observed high and ultra-high energy CRs, as well as neutrinos through particle cascades \citep[e.g.,][]{mannheim1994interactions,aharonian2002proton}.

The modeling of either of these radiative processes requires knowledge of the geometrical structure of the emitting region. Observations show jets that remain collimated up to large distances, following cylindrical or conical structures \citep[e.g.,][]{lister2013mojave,hada2016high}. However, for simplicity, spectral models often consider localized and spherical single-zone accelerating regions because they provide a good first-order approximation \citep[e.g.,][]{tavecchio1998constraints,mastichiadis_kirk_2002,marscher2008inner}. In order to correctly factor in the observed jet geometry, we need to describe an accelerating and expanding outflow, and properly connect its physical properties with those of the accretion flow. Such inhomogeneous multi-zone jet models are able to self-consistently produce both the characteristic flat-to-inverted radio spectra observed in many compact jet systems, and the upscattered high-energy continuum \citep{blandford1979relativistic,hjellming1988radio}.

Multiple groups have considered such multi-zone models in the past. For instance, \cite{falcke1995jet} derived a simple model for the dynamical properties of a hydrodynamically driven, self-collimating jet, assumed to be powered by the accretion flow. This model was further developed with jet-intrinsic particle distributions and more detailed radiative calculations, and extended to XRBs by \cite{markoff2001jet} and \cite{Markoff2005}. The semi-analytical nature of this model has the great advantage that one can directly fit its physical parameters to data. Numerical simulations of the detailed magnetohydrodynamics of the jet flow, combined with radiative transfer calculations, would be very computational expensive and time consuming for such a task.

In this work, we adopt the multi-zone leptonic model of \cite{Markoff2005} in its most recent version  (\citealt{maitra2009constraining,crumley2017symbiosis,lucchini2019breaking}; Lucchini et al. submitted) and we further develop it by including hadronic interactions. This is the first hadronic multi-zone jet model for Galactic sources that additionally includes further improvements to the already implemented leptonic ones, such as pair cascades \citep{coppi1990reaction,bottcher1997pair}. 

An ideal source to test our newly developed model, is one of the brightest and well-studied black-hole high-mass XRB, Cyg~X--1 and its persistent jets \citep{stirling2001relativistic,rushton2012weak}. Along with the model, we present a new data set obtained by the CHOCBOX campaign (Cyg X-1 Hard state Observations of a Complete Binary Orbit in X-rays: \citealt{uttley2017}). This campaign performed simultaneous observations with the satellite observatories \textit{XMM-Newton}, \textit{NuSTAR}, and \textit{INTEGRAL}, which, together with the ground-based interferometers (NOEMA, VLA, and VLBA) provide the first multi-wavelength data set of that kind for Cyg~X--1.

We also include the most recent X-ray polarisation information for Cyg~X--1. Linear polarisation has been reported in the energy band below 200\,keV but the polarisation fraction is strongly energy-dependent and does not exceed 10 per cent \citep{chauvin2018polarization,Chauvin2018pogo}. In contrast, the hard X-ray emission in the 0.4--2\,MeV band is linearly polarised at a level of $\sim$70 per cent \citep{2011Sci...332..438L,jourdain2012separation,0004-637X-807-1-17}. Such a high polarisation fraction can only be explained as synchrotron emission from an ordered magnetic field, and places strong constraints on the modelling. In this work, we assume that the synchrotron radiation originates in the compact jets of Cyg~X--1.

For this work, we adopt the updated distance and black-hole mass for Cyg~X--1 of $2.22\,$kpc and $21.4 \,\rm{M_{\odot}}$, respectively (Miller-Jones et al. subm). The distance is in good agreement with the \textit{Gaia} DR2 distance of $2.38^{+0.20}_{-0.17}$ \citep{brown2018gaia}, which is about 30 per cent more distant than previously thought \citep{reid2011trigonometric}. The mass of the black hole was historically estimated to be between $14.8 \,\rm{M_{\odot}}$ \citep{orosz2011mass} and $16 \,\rm{M_{\odot}}$ \citep{2014MNRAS.440L..61Z,mastroserio2019x}, significantly lower than the updated value. The impact of the updated value of the mass of the black hole can be significant making the revision of modeling the source necessary. The jet inclination angle is 27.5$^{\circ}$.  The companion is a $\sim41$M$_\odot$ star (Miller-Jones et al. subm), which is about twice as massive as the foregoing estimate by \cite{orosz2011mass}. The spectral type of the companion star is O9.7 Iab \citep{1972Natur.235..271B}. The binary separation is estimated to be $\sim 3.7\times 10^{12}\,$cm \citep{miller2005revealing} and the system orbital period is around 5.6 days \citep{1972Natur.235...37W}.

This paper is organized as follows. We discuss the new observational data set of Cyg~X--1 in Section \ref{section: observational data} and our new lepto-hadronic model in Section \ref{section: model}. In Section \ref{section: results} we present the results of our modelling. Finally, we outline in Section \ref{section: discussion} the significance of the results and summarize our work in Section \ref{section: summary}.

\section{Observations and Data Extraction}\label{section: observational data}

The bulk of the data we use to constrain the physical parameters of our model resulted from the CHOCBOX campaign \citep{uttley2017}. In particular, we select data within the time interval 2016 May 31 05:15:01.5 -- 07:07:04.5 UTC, which provides simultaneous coverage by NOEMA, \xmm , \nustar , and \integral .

In addition, we consider some supplemental, non-simultaneous, long-term averaged archival data. We use the mid-infrared data \citep{rahoui2011multiwavelength} to constrain physical properties of the donor star. We take into account a long-term 15-year average MeV spectrum by \integral\ \citep{Cangemi2020longterm} as well as the publicly available GeV $\g$-ray spectrum from the \textit{Fermi/LAT} collaboration \citep{zanin2016detection}. The low flux and challenging detection techniques require averaging the data over longer timescales. \cite{Cangemi2020longterm} are the first to average over all existing \integral\ data of Cyg~X--1 in its hard state. The $\g$-ray spectrum we use here comprises data averaged over 7.5\,years, only during the hard state of Cyg~X--1. Averaging thus provides the best-possible constraints to the MeV and GeV emission at the moment. While modeling, we do take into account the systematics arising from integrating over flux variations. We list all the data we use in this work in Table \ref{table: data}.

\subsection{Very Large Array (VLA)}

We observed Cyg~X--1 with the Karl G.\ Jansky Very Large Array (VLA) on 2016 May 31, from 04:29--08:28 UT, under project code VLA/15B-236.  The VLA observed in two subarrays, of 14 and 13 antennas spread approximately evenly over each of the three arms of the array, which was in its moderately-extended B configuration.  The first subarray observed primarily in the Q-band, with two 1024-MHz basebands centred at 40.5 and 46.0\,GHz, and the second observed primarily in the K-band, with the two 1024-MHz basebands centred at 20.9 and 25.8\,GHz.  Each subarray observed a single two-minute scan at a lower frequency (two 1024-MHz basebands centred at 5.25 and 7.45\,GHz, and a single 1024-MHz baseband centred at 1.5\,GHz, respectively) to characterise the broadband spectral behaviour.  We used 3C\,286 as the bandpass and delay calibrator, and to set the flux density scale, and we derived the complex gain solutions using the nearby extragalactic source J2015+3710.

We processed the data using the Common Astronomy Software Application \citep[CASA;][]{McMullin2007}.  The data were initially calibrated using the VLA CASA Calibration Pipeline (v4.5.3), and after some additional flagging to excise radio frequency interference, we imaged the target data using CASA version 4.5.2.  The low elevation at the beginning of the run caused significant phase decorrelation and an increased system temperature.  Although we were able to self-calibrate the data in phase down to a solution timescale of 2 minutes, the flux densities were still found to be biased low. We therefore restricted our images to the final 90\,min of the run. Cygnus X--1 was significantly detected in all images, which were made with Briggs weighting, with a robust parameter of 1.

\subsection{NOrthern Extended Millimeter Array (NOEMA)}

The NOEMA observations of Cyg~X--1 (project code: W15BQ, PI: Tetarenko) took place on 2016 May 31 (05:15:01-07:52:53.0 UT, MJD 57539.2188 - 57539.3284), in the 2\,mm (tuning frequency of 140\,GHz) band. These observations were made with the WideX correlator, to yield 1 base-band, with a total bandwidth of 3.6 GHz per polarisation. The array was in the 6ant-Special configuration (N02W12E04N11E10N07), with 6 antennas, spending 1.9 hrs on source during our observations. We used J2013+370 as a phase calibrator, 3C454.3 as a bandpass calibrator, and MWC349 as a flux calibrator. We performed phase only self-calibration on the data, with a solution interval of 45 seconds. The weather significantly degraded after 07:07 UT at NOEMA, therefore we do not include data after that time in our analysis. As CASA is unable to handle NOEMA data in its original format, flagging and calibration of the data were first performed in \textsc{gildas}\footnote{\href{http://www.iram.fr/IRAMFR/GILDAS}{http://www.iram.fr/IRAMFR/GILDAS}} using standard procedures, then the data were exported to CASA\footnote{To convert a NOEMA data set for use in CASA, we followed the procedures outlined at \href{https://www.iram.fr/IRAMFR/ARC/ documents/filler/casa- gildas.pdf}{https://www.iram.fr/IRAMFR/ARC/ documents/filler/casa- gildas.pdf}.} for imaging (with natural weighting to maximize sensitivity). The flux density of the source was measured by fitting a point source in the image plane (using the \texttt{imfit} task). 

\subsection{\xmm }
We consider the \xmm\ observation ID 0745250501, which observed Cyg~X--1 in timing mode using its EPIC-pn camera \citep{Strueder2001} for a total of about 145\,ks. First, we create calibrated and filtered event lists using the \texttt{SAS~v.16.1.0}, which we further correct for X-ray loading and flag soft flare events. We consider only counts strictly simultaneous to the NOEMA observation time period resulting in a net exposure time of 3.5\,ks. We use the filtered event lists to extract 0.3--10\,keV spectra according to standard procedures.

\subsection{\nustar }
\nustar\ \citep{Harrison2013} measures photons up to $\sim 80\,$keV by focusing hard X-rays on two focal-plane modules FPM~A and FPM~B. We extract data from within 3--78\,keV with the standard \nustar\ Data Analysis Software \texttt{NuSTARDAS-v.1.8.0} as part of \texttt{HEASOFT-v.6.22.1}. Due to the high flux of Cyg~X--1, we extract source counts from within a relatively large region of 150\arcsec\ radius on both chips FPM~A and FPM~B, and background counts from a region of 100\arcsec\ located off-source but close enough not prevent bias due to the spatial background dependence \citep{Wik2014}. To make sure to have simultaneous coverage with the observational time window of NOEMA, we define appropriate good-time intervals for the observation ID 30002150004, which results in a net exposure time of 1.9\,ks each for FPM~A and FPM~B.

\subsection{\integral}
We extract the \integral\ Soft Gamma-Ray Imager (ISGRI; \citealt{lebrun2003}) data with the \texttt{Off-line Scientific Analysis (OSA)} software v10.2 to match the simultaneous time interval as much as possible, resulting in the use of three science windows, 168500020010, 168500030010 and 168500040010 and 6.5\,ks effective exposure time.

The state-resolved scientific products (images, light curves, and spectra) of the coded-mask instrument ISGRI were obtained with standard procedures. We extract spectra and images of Cyg X-1 on a single-science-window (scw) basis. For each scw, we construct a sky model including the brightest sources active in the field at the time of observation as found from the analysis of the full CHOCBOX \integral exposure, i.e. Cyg~X--1, Cyg~X--3, Cyg A, GRO\,J2058$+$42, KS\,1947$+$300 and SAX J2103.5+4545.

%%%%%%%%%%%%%%%%%%%%%%%%%%%%%%%
%%%% DATA TABLE %%%%
%%%%%%%%%%%%%%%%%%%%%%%%%%%%%%%
\begin{table*}
	\begin{center}
		\setlength{\tabcolsep}{6pt} % Default value: 6pt
		\renewcommand{\arraystretch}{1.4} % Default value: 1
		\begin{tabular}[b]{lcccc}\hline\hline
			Observatory & log Frequency (Hz) & log Energy (eV) & Flux Density (mJy$^{\rm{a}}$)& References\\ \hline
		    VLA& $\begin{array}{cc}
			      10.32 & 10.41\\
			      10.61  & 10.66
			\end{array}$&
			$\begin{array}{cc}
			      -4.07 & -3.97\\
			      -3.78  & -3.72
			\end{array}$
		     &
			$\begin{array}{cc}
			      8.07\pm 0.03 & 8.11 \pm 0.03\\
			      8.66 \pm 0.10  & 8.14\pm 0.15
			\end{array}$&This work\\
			NOEMA& 11.15& $-3.24$ &$6.87\pm 0.27$& This work\\
			\textit{Spitzer} &12.97--13.77&$-1.42$-- $-0.61$& $54.57$ at $10^{13}\,$Hz & \cite{rahoui2011multiwavelength} \\
			\xmm & 16.86--18.38 & 2.48--4.0 & $\begin{array}{c}
			     0.07~ \rm{at}~ 3\,\rm{keV}  \\
			     0.32~ \rm{at}~ 10\,\rm{keV}
			\end{array}$ & This work\\
			\nustar & 17.87--19.28 & 3.49--4.89 & $\begin{array}{c}
			     0.54~\rm{at}~ 3\,\rm{keV}  \\
			     0.18~\rm{at}~78\,\rm{keV} 
			\end{array}$ & \multirow{1}{*}{This work}\\
			\multirow{2}{*}{\integral} &%$\begin{array}{c}
			     18.78--19.68 & 4.40--5.30 &0.19 at 25\,keV,~ 0.02 at 200\,keV & This work\\ 
            & 19.73--20.90 & 5.35--6.52&  0.01 at 225\,keV, $10^{-4}$ at~ 3.3\,MeV&\cite{Cangemi2020longterm}
			\\
			\textit{Fermi/LAT}&22.43--24.43&8.05--10.05&$\begin{array}{c}
			        7\times 10^{-5}~\rm{at}~ 0.1\,\rm{GeV}\\
			        2\times 10^{-9}~\rm{at}~ 10\,\rm{GeV}
			\end{array}$ & \cite{zanin2016detection}\\ \hline
		\end{tabular} 
		\caption{The observational multiwavelength data used in this work. $^{\rm{a}}$mJy$\,= 10^{-26}\,\rm{erg\,cm^{-2}\,s^{-1}\,Hz^{-1}}$. }
		\label{table: data}
	\end{center}
\end{table*}

\section{Model Details}\label{section: model}
\subsection{Dynamical Quantities}\label{section: dynamics}

We  describe the multi-zone jet model based on \cite{Markoff2005} and its extensions referenced above. In this section we summarize the major properties of the model and focus on our new extension of including the effect of hadronic particle acceleration and secondary production.

A fully self-consistent jet model should solve the force balance equations along the streamlines and perpendicular to them. This calculation would yield the radial profile and the acceleration profile describing a given jet configuration starting from a set of initial conditions.  For simplicity we assume a fixed shape for the jet radial profile, based on observational evidence in AGN, which together with the longitudinal velocity profile then determines the profiles along the jet of the number density, and global magnetic field strength. Specifically, the cross-sectional radius $R$ at any height $z$ along the jet is given by
\begin{equation}\label{radius profile}
 R\left( z\right)  =  R_0 + \left(z - z_0\right) \frac{\Gamma_0\beta_0}{\Gamma_j\beta_j},
\end{equation}
where $R_0$ is the radius of the jet base, $z_0$ is the height of the jet base above the black hole, $\beta_{0,j}$ and $\beta_j$ are the bulk velocity of the plasma at the jet base and at height $z$ respectively, and $\Gamma$ is the corresponding Lorentz factor. 

The solution of the Euler equation \citep{crumley2017symbiosis}
\begin{equation}
  \begin{split}
    \left\{ \Gamma_j\beta_j\frac{\Gamma_{\rm{ad}}+\xi}{\Gamma_{\rm{ad}}-1} - \Gamma_{\rm{ad}}\Gamma_j\beta_j - \frac{\Gamma_{\rm{ad}}}{\Gamma_j\beta_j}+\frac{ 2(z-z_0)\Gamma_0\beta_0/(\Gamma_j\beta_j)}{R_0\Gamma_j\beta_j+\Gamma_0\beta_0(z-z_0)} \right\} & \\ \times \frac{\partial \Gamma_j\beta_j}{\partial z} = \frac{2\Gamma_0\beta_0}{R _0\Gamma_j\beta_j + \Gamma_0\beta_0(z-z_0)} &
    \end{split}
\end{equation}
gives the velocity profile along the jet $\Gamma_j(z)$. In the above equation, $\Gamma_{\rm{ad}}$ is the adiabatic index of the flow (5/3 for a non-relativistic and 4/3 for a relativistic flow), 
\begin{equation}
\xi = \left( \frac{\Gamma_j\beta_j}{\Gamma_0\beta_0} \right)^{\Gamma_{\rm{ad}}-1};\, \Gamma_0\beta_0 = \sqrt{ \frac{ \Gamma_{\rm{ad}} (\Gamma_{\rm{ad}}-1)
} {1+2\Gamma_{\rm{ad}}-\Gamma_{\rm{ad}}^2}.
}
\end{equation}

Conservation of the particle number density results in;
\begin{equation}
 n\left(z\right) = n_0 \left( \frac{\Gamma_j\beta_j}{\Gamma_0\beta_0} \right)^{-1} \left(\frac{R}{R_0}\right)^{-2},
\end{equation}
where $n_0$ is the differential number density at the jet base in $\rm{cm^{-3}\,erg^{-1}}$. For a quasi-isothermal jet, which seems to be necessary to explain the flat/inverted spectrum, the internal energy density is given by (see \citealt{crumley2017symbiosis}):
\begin{equation}
 U_j\left(z\right) = n_0\rm{m_pc}^2 \left( \frac{\Gamma_j\beta_j}{\Gamma_0\beta_0} \right)^{-\Gamma_{\rm{ad}}} \left(\frac{R}{R_0}\right)^{-2},
\end{equation}
where $\rm{m_{p}c}^2$ is the rest-frame energy of the protons that carry most of the kinetic energy. By assuming a fixed plasma beta parameter $\beta = U_\mathrm{e}/U_\mathrm{B}$, where $U_\mathrm{e}$ is the internal energy density of the electrons, and $U_\mathrm{B}$ the magnetic energy density, we can determine the profile of the magnetic field along the jet to be
\begin{equation}\label{B field profile}
 B\left(z\right) = \sqrt{\frac{8\pi U_\mathrm{e}\left(z\right)}{\beta}},
\end{equation}
where the energy density of the magnetic field is $U_\mathrm{B} = B^2/8\pi$. For simplicity, we do not distinguish between toroidal and poloidal components but we assume that the field is tangled with a characteristic strength.

In addition to the jets, which include a thermal-dominated, corona-like region at their base, we incorporate a simple description for an additional thermal compact corona located around the black hole. We assume that a hot electron plasma of temperature $T_{\rm{cor}}$ is covering a radius $R_{\rm{cor}}$ and has an optical depth $\tau_{\rm{cor}}$. These hot electrons inverse Compton upscatter the black body photons emitted by the accretion disc, while the thermal population in the jet base can upscatter both disc photons as well as synchrotron photons.

\subsection{Particle distributions}

Thermal electrons\footnote{We do not distinguish between electrons and positrons. The results in this work do not depend on the charge of the lepton.} are assumed to be directly injected into the jet base from the accreting inflow with a thermal Maxwell-J\"{u}ttner distribution, which reduces to the standard Maxwellian form in the non-relativistic case. Protons can be found in the jet base as well but they are entirely cold, and only carry the kinetic energy of the jet. The initial number density of the protons carried by the jet is defined as
\begin{equation}\label{cold protons distribution}
 n_{0} = \frac{L_{\rm{jet}}}{4\beta_{0,s}\Gamma_{0,s}c\,\rm{m_pc}^2\pi R_0^2}, 
\end{equation}
where half of the injected power $L_{\rm{jet}}$ goes into cold protons, while the other half is shared by the magnetic field and leptons, thus the factor 1/4. We assume equal number density of electrons and protons. Further, $\beta_{0,s}\Gamma_{0,s}c$ is the sound speed of a relativistic fluid with adiabatic index 4/3. The total injected power $L_{\rm jet}$  is a free parameter of the model and is assumed to be proportional to the accretion energy $\dot{M}c^2$.

Once the particles propagate out some distance $z_{\rm{diss}}$ along the jet, a fitted parameter, we assume that a fixed fraction (10 per cent) of both leptons and hadrons are accelerated into a power-law with index $p$ from this point onwards. We do not invoke any particular acceleration mechanism nor distinguish between acceleration or re-acceleration. We thus allow the power-law index $p$ to be a free parameter in our model. Moreover, we assume constant particle acceleration beyond the particle acceleration region $z_{\rm{diss}}$. Another free parameter is the acceleration efficiency $f_{\rm{sc}}$ \citep[see e.g.,][]{jokipii1987rate,aharonian2004very}. Given this efficiency, the maximum energy achieved by the particles is calculated self-consistently along the jet by considering the main physical processes that limit the further acceleration of particles. The dominant cooling mechanisms are synchrotron radiation and inverse Compton scattering (ICS) for leptons, and escape from the source for hadrons. Adiabatic cooling is not relevant because the jets are actively collimated.

In order to calculate the particle distributions along the jets, we solve the continuity equation, which in energy phase space can be written in the general form:
\begin{equation}\label{general kinetic equation}
  \begin{split}
    \frac{\partial N_i\left( E_i,t,z\right) }{\partial t} + \frac{\partial \left( \Gamma_{j}v_{j}N\left( E_i,t,z\right) \right) }{\partial z} & \\  + 
    \frac{\partial \left( b\left( E_i,t,z\right) N_i\left( E_i,t,z\right) \right) }{\partial E_i}  -\frac{N_i\left( E_i,t,z\right) }{\tau_{\rm{esc}}\left( E_i,t,z\right) }    
      =  Q\left( E_i,t,z\right). & 
  \end{split}
\end{equation}
The above equation describes the temporal evolution of the number density of the particle population $i$, i.e. electrons or protons. Since we assume a steady-state source, we neglect the first term on the left-hand side, making every quantity time-independent. We also neglect the effects of spallation and diffusion.

The second term on the left-hand side describes the propagation of particles along the jet. The third term expresses the radiative cooling of the particles, i.e. synchrotron radiation and ICS for leptons, as well as inelastic collisions for hadrons. The particles may escape the source within the timescale $\tau_{\rm{esc}}\left(E_i,t,z\right)$, which in our treatment is only energy-dependent. Finally, the right-hand side describes the injection term, which is the sum of a Maxwell-J\"{u}ttner thermal distribution at low energies and a non-thermal power-law  with an exponential cutoff at the self-consistently derived maximum energy. The non-thermal power-law is included only starting at the dissipation region $z_{\rm{diss}}$ where particle acceleration initiates. 

Losses will dominate over acceleration above some particular energy $E_{\rm{max}}$ which can be self-consistently calculated -- here for the leptonic case -- by setting
\begin{equation}\label{electron max energy}
  \tau_{\rm{acc}}^{-1}\left( E_{e,\rm{max}}\right)  =  \tau_{\rm{syn}}^{-1} \left( E_{e,\rm{max}} \right)  + \tau_{\rm{ICS}}^{-1}\left( E_{e,\rm{max}}\right)  + \tau_{\rm{esc}}^{-1}\left( E_{e,\rm{max}}\right), 
\end{equation}
with the timescales for acceleration, synchrotron cooling, ICS cooling in the Thomson regime, and the escape of leptons, i.e.
\begin{itemize}
  \item $\tau_{\rm{acc}} = \dfrac{4E_e}{3f_{\rm{sc}} \rm{ec}B}$
  \item $\tau_{\rm{syn}} = \dfrac{6\pi \rm{m_e}^2 \rm{c}^3}{\sigma_{\rm{T}} B^2 E_e\beta_e^2}$
  \item $\tau_{\rm{ICS}} = \tau_{\rm{syn}}\dfrac{U_B}{u_{\rm{rad}}}$
  \item $\tau_{\rm{esc}} = \dfrac{R}{\beta_e \rm{c}}$, 
\end{itemize}
respectively. Here, e is the electron charge, $B$ the magnetic field of the jet at height $z$ with radius $R$, $\rm{m_e}$ the rest mass of the electron, c the speed of light, $\sigma_{\rm{T}}$ the Thomson cross-section, $\beta_e$ the speed of the particle in units of c, $U_{\rm{B}}=B^2/8\pi$ the energy density of the magnetic field, $u_{\rm{rad}}$ the energy density of the radiation field upscattered by the electrons.

Following the same approach, we calculate the maximum energy of protons in case of hadronic acceleration by setting
%\begin{eqnarray}
    %\centering
    \begin{multline}\label{proton max energy}
        \tau_{\rm{acc}}^{-1}\left(E_{p,\rm{max}}\right) =\,  \tau_{\rm{syn}}^{-1}\left(E_{p,\rm{max}}\right) + \tau_{\rm{pp}}^{-1}\left(E_{p,\rm{max}}\right) +\\     
         \tau_{\rm{p\gamma}}^{-1}\left(E_{p,\rm{max}}\right) + \tau_{\rm{esc}}^{-1}\left(E_{p,\rm{max}}\right),
    \end{multline}
%\end{eqnarray}
with the timescales for acceleration, synchrotron cooling, proton-proton collisions, proton-photon collisions, and the escape of protons, i.e.
\begin{itemize}
  \item $\tau_{\rm{acc}} = \dfrac{4E_p}{3f_{\rm{sc}} \rm{ec}B}$
  \item $\tau_{\rm{syn}} = \dfrac{6\pi \rm{m_p}^2 \rm{c}^3}{\sigma_{\rm{T}} B^2 E_p\beta_p^2}\times \left( \rm{\dfrac{m_p}{m_e}}\right)^2$
  \item $\tau_{\rm{pp}}  = \left( K_{\rm{pp}}\sigma_{\rm{pp}}n_{\rm{th}}c\right) ^{-1}$
  \item $\tau_{{p\g}} =  \left( K_{\rm{p\g}}\sigma_{\rm{p\g}}n_{\g}c\right) ^{-1}$
  \item $\tau_{\rm{esc}} = \dfrac{R}{\beta_p c}$ 
\end{itemize}
Here, $K_{\rm{pp}}$ corresponds to the multiplicity (average number of secondary particles), $\sigma_{\rm{pp}}$ to the cross-section of this interaction, and $n_{\rm{th}}$ to the number density of the target particles (see Section \ref{hadronic processes}). For proton-photon interactions between the accelerated protons and a photon field with number density $n_{\g}$, we consider the multiplicity $K_{\rm{p\gamma}}$ \citep{mannheim1994interactions}. One can see that the proton-synchrotron timescale is approximately $\left(\rm{m_p/m_e}\right)^3$ times longer than the electron one. 

The injection term becomes a power-law with an exponential cutoff beyond the particle acceleration region $z_{\rm diss}$, i.e.
\begin{equation}\label{injection term}
  Q\left(E_i\right) = Q_0 E_i^{-p}\times \exp{\left(-E_i/E_{i,\rm{max}}\right)},
\end{equation}
where $Q_0$ is a normalisation factor and $p>0$ is allowed to vary between 1.5 and 2.5, consistent with standard particle acceleration mechanisms. The power-law index is assumed to be equal for electrons and protons, which implies a common acceleration mechanism for both populations. Equation \ref{injection term} is the less computationally-expensive form of the output of Particle-In-Cell (PIC) simulations where the thermal particle distribution leads to a self-consistent formation of a power-law of accelerated particles in time \citep[e.g.,][and references therein]{sironi2009particle,crumley2019kinetic}. We include further distributions of secondary pairs from hadronic processes and photon-photon annihilation (see below) into this injection term $Q$.  

\subsection{Radiative Processes}\label{section: radiative processes}
\subsubsection{Leptonic Processes}
Electrons throughout the jet lose energy due to synchrotron and IC radiation. Before the particle acceleration region, even thermal electrons emit synchrotron radiation due to the relatively strong magnetic field. Beyond the particle acceleration region, the non-thermal leptonic process that dominates is the synchrotron radiation. For electron ICS we include photon fields from synchrotron radiation (synchrotron-self Compton -- SSC), the disc around the black hole, and the companion star. We take into account the geometry of the companion star because, for high-mass XRBs like Cyg~X--1, the size of the star is comparable to the size of the jet, especially for regions close to the compact object where the majority of the high energy radiation is likely to originate. In particular, we calculate the photon field of the companion star as seen in the jet frame accounting for the Doppler boosting (each jet segment travels at a different Lorentz factor). All expressions for synchrotron radiation and ICS are taken from \cite{blumenthal1970bremsstrahlung} and \cite{rybicki2008radiative}.

Furthermore, we include the full treatment of photon-photon annihilation and electromagnetic cascades \citep{coppi1990reaction,bottcher1997pair}. Depending on the number density of produced pairs, additional interactions between electrons and positrons can cause pair-annihilation leading to the production of $\g$-rays. This process can occur until the lepton energy budget becomes insufficient for further photon production. The photon fields we take into account are the same as for ICS. Finally, we add the produced pairs to the leptonic population, which are then cooled as described above.

\subsubsection{Hadronic Processes}\label{hadronic processes}
In the case where protons and/or ions are accelerated to relativistic energies in the jet, they can inelastically collide with thermal protons and photons inside the jet flow and produce secondary particles \citep{mannheim1994interactions}. In the extension of our model, we therefore implement both proton-proton and proton-photon interactions. We use the full semi-analytical treatment of \cite{kelner2006energy} and \cite{kelner2008energy} based on Monte-Carlo simulations (see below for more details).   

\paragraph{Proton-proton interactions\\}
Collisions of non-thermal protons with thermal jet protons and stellar-wind protons  (proton-proton collisions, pp, henceforth) lead to the production of $\g$-rays, secondary electrons, and neutrinos. The interactions responsible for the production of these particles can be described as
\begin{eqnarray*}
  \rm{p + p  \rightarrow  p + p + \alpha \pi^0 + \beta \left(\pi^+ + \pi^-\right)},
\end{eqnarray*}
where $\rm{\alpha}$ and $\rm{\beta}$ are the collision energy-dependent multiplicity of the related products \citep[see e.g.,][]{Romero2017}. The charged pions decay as
\begin{eqnarray*}
  \rm{\pi^+ \rightarrow \mu^+ + \nu_{\mu}}, & &         ~\rm{\mu^+ \rightarrow e^+ + \nu_e + \bar{\nu}_{\mu}}, \\
  \rm{\pi^- \rightarrow \mu^- + \bar{\nu}_{\mu}}, & &   ~\rm{\mu^- \rightarrow e^- + \bar{\nu}_e + \nu_{\mu}},
\end{eqnarray*}
and the neutral pions decay into two gamma-rays, i.e.
\begin{eqnarray*}
  \rm{\pi^0 \rightarrow \gamma + \gamma}.
\end{eqnarray*}

In order for these interactions to occur, the energy of the accelerated proton has to exceed the threshold of $E_{\rm{th}} \simeq 1.22\,$GeV \citep{mannheim1994interactions}.  

The lifetime of the produced mesons is well measured by laboratory experiments and short compared to the dynamical timescales of the jet. We can therefore assume instant decays. Consequently, the charged products do not radiatively lose energy as they would in extreme environments of either very strong magnetic fields or very high energies \citep[e.g.,][]{MUCKE2003protonBLLac}. The above statement can be parametrized as follows \citep[e.g.,][]{boettcher2013leptohadronic}
\begin{equation}
    \rm{B\gamma_p}\ll 
    \begin{cases}
    7.8\times 10^{11} \rm{\,G~ for~pions} \\
    5.6\times 10^{10} \rm{\,G~ for~muons,} 
    \end{cases}
\end{equation}
where B is the strength of the magnetic field in the jet rest frame and $\g_p$ the Lorentz factor of the proton. Given that the highest value of the magnetic field is in the jet base ($10^7\,$G) and that hadronic interactions do not occur yet because particle acceleration occurs later, one can see that the above inequality is always satisfied. 

In order to produce the distributions of stable products, we follow the semi-analytical approximation of \cite{kelner2006energy}. In particular, the differential number density of the $\g$-rays is given by the expression:
\begin{equation}\label{gamma rays from pp neutral pion decay}
 \frac{dn_{\gamma}\left(z,E_{\gamma}\right)}{dE_{\gamma}} = cn_{\rm{targ}}\int_{0}^{1} \sigma_{\rm{pp}}\left(\frac{E_{\gamma}}{x}\right)n_p\left(z,\frac{E_{\gamma}}{x}\right) F_{\gamma}\left(x,\frac{E_{\gamma}}{x}\right)\frac{dx}{x},
\end{equation}
where $E_{\g}$ is the energy of the $\g$-ray, $n_{\rm{targ}}$ is the number density of the thermal target protons, $\sigma_{\rm{pp}}$ is the cross section for pp collisions, $\rm{n_p}$ is the number density of the non-thermal protons, $\rm{x = E_{\g}/E_p}$ is the normalized photon energy with respect to initial proton energy and $\rm{F_{\g}\left(x,E_{\g}/x\right)}$ is the spectrum of $\g$-rays.

The cross section for pp interactions can be given by the semi-analytical expression
\begin{equation}\label{cross section pp}
\begin{split}
    \sigma_{\rm{pp}}\left(T_p\right) = \left[ 30.7 - 0.96\log \left( \frac{T_p}{T_{\rm{thr}}} \right) + 0.18\log ^2 \left( \frac{T_p}{T_{\rm{thr}}} \right) \right]\\ \times \left[ 1 - \left( \frac{T_{\rm{thr}}}{T_p} \right)^{1.9} \right]^3\,\rm{mb},
\end{split}
\end{equation}
where $T_p$ is the proton kinetic energy in the laboratory frame and $T_{\rm{thr}} =2m_{\pi}+m_{\pi}^2/2m_p \simeq 0.2797\,$GeV the threshold kinetic energy for this interaction to take place \citep{kafexhiu2014parametrization}. \cite{kelner2006energy} provide semi-analytical calculations for the $\g$-ray spectrum as well as the other secondary particles. 

For this work the target protons are the cold protons of the jet and protons emitted by the heavy companion star in the form of a homogeneous stellar wind. In particular, the companion star of Cyg~X--1 is a blue supergiant that loses $\sim 10^{-6}$ M$_\odot$/yr in the form of stellar wind \citep{Gies_2008}. We use the following expression to calculate the proton number density emitted by the companion
\begin{equation}\label{stellar wind}
 n_{\rm{wind}}\left(z\right) = \frac{\dot{M}_{\star}}{4\pi\left(\alpha^2_{\star} + z^2\right) v_{\rm{wind}}{\rm{m_p}}} \times \left[ 1 - \frac{R_{\star}}{\sqrt{\alpha^2_{\star} + z^2}}  \right]^{-\beta_{\rm{wind}}}
\end{equation}
\citep{Grinberg2015variability}, where $\dot{M}_{\star} = 4\pi \rho\left(r\right) v\left(r\right)$ is the mass-loss rate based on the radially-dependent mass density profile $\rho(z)$, $v_{\rm{wind}}$ is the terminal velocity of the wind on the jet wall, $\alpha^2_{\star}$ is the distance of the massive star from the black hole, $R_{\star}$ is the radius of the massive star, $z$ is the distance from the central black hole along the jet axis and $\beta_{\rm{wind}}$ is a free parameter used to improve the velocity profile of the wind found to be 1.6 \citep[see e.g.,][]{Grinberg2015variability}. From geometrical, filling-factor considerations, we assume that only 10 per cent of the wind protons take part in the pp process \citep[see e.g.,][]{pepe2015lepto}. Therefore, the total target number density (in cm$^{-3}$) is given by:
\begin{equation}
  n_{\rm{targ}}\left(z\right) = 0.1n_{\rm{wind}}\left(z\right) + n_{\rm{p},\rm{cold}}\left(z\right).
\end{equation}

\paragraph{Proton-photon interactions\\}
In addition to the pp interaction, inelastic collisions between non-thermal protons and photons occur in the jet (p$\g$ henceforth). For this process we take into account the same photons fields as described above for leptonic ICS.  

Depending on the centre-of-mass energy of the inelastic collision, we consider two processes: photopair and photomeson interactions. The photopair interaction is a p$\g$ collision resulting in the production of an electron-positron pair
\begin{eqnarray*}
 \rm{p + \gamma \rightarrow p + e^+ + e^-},
\end{eqnarray*}
also called the Bethe-Heitler process.
Alternatively, a p$\g$ collision can result in the production of mesons, similarly to the pp interaction discussed above. The photomeson process can be written as
\begin{eqnarray*}
  \rm{p + \g  \rightarrow   p + p + \alpha \pi^0 + \beta \left(\pi^+ + \pi^-\right)}.
\end{eqnarray*}

The energy thresholds for photopair and photomeson processes to occur are:
\begin{eqnarray}
  E_{p,\rm{thres}} &=& \rm{4.8\times 10^{14}/\epsilon_{\rm{eV}}~eV ~for ~photopair},\\
  E_{p,\rm{thres}}&=& \rm{7.1\times 10^{16}/\epsilon_{\rm{eV}}~eV ~for ~photomeson},
\end{eqnarray}
where $\epsilon_{\rm{eV}}$ is the energy of the target photon in eV. The photopair process has a lower energy threshold to occur. However, if the energy threshold for the photomeson process is met, then the energy loss of the proton is more significant compared to the photopair process, making the photomeson process dominant \citep{mannheim1994interactions}.

Semi-analytical expressions for the distributions of stable secondary particles are provided by \cite{kelner2008energy}. Secondary particles produced in the above processes can further interact within the jet before escaping. In this paper we do not add the secondary leptons to the primary leptonic population, but rather calculate their radiative processes and their relative contribution to the electromagnetic spectrum separately, for comparison.

\subsection{Corona model}\label{section: corona}
Along with the jet, we include an additional component in the form of a simple spherical corona surrounding the accretion disc. As discussed in section \ref{section: best fits}, this is necessary in order to match the X-ray emission of the source.

We assume that the electrons in the corona are thermal with a temperature $T_{\rm{cor}}$, and that the entire corona is described by an optical depth $\tau_{\rm cor}$ and a radius $R_{\rm cor}$. We define the number density of the injected electrons as: $n_{\rm {e,cor}} = \tau_{\rm {cor}}/\sigma_{\rm{ T}}R_{\rm {cor}}$, where $\sigma_{\rm T}$ is the Thomson cross section. For the emission related to the corona, we only consider the disc photons as the source of seed photons for ICS, and we calculate the radiation energy density of the seed photons at the centre of the system. This means that the coronal radius $R_{\rm cor}$ effectively acts as a normalisation constant, rather than representing the exact physical radius of the X-ray emitting region.

\section{Results}\label{section: results}
We perform simultaneous spectral fits of all data presented in Section \ref{section: observational data} using the Interactive Spectral Interpretation System (\texttt{ISIS}; \citealt{houck2000ISIS}). We explore the parameter space using a Markov Chain Monte Carlo (MCMC) method and its implementation via the \texttt{emcee} algorithm. In particular, we initiate 20 walkers per free parameter and perform $\sim10^4$ loops. The chains require a significant number of loops before they successfully converge, so we exclude the 50 per cent of the initial loops. We use the rest of the loops to derive the uncertainties of each free parameter (shown in Table~\ref{table: parameters for models}. The fixed parameters including those of the donor star as assumed by \cite{Grinberg2015variability} are given in Table~\ref{table:common parameters}. 
 The free parameters we allow to vary during the fitting are shown in Table \ref{table: parameters for models}. These are the injected power to the jet base $L_{\rm{jet}}$, the radius of the jet base $R_0$, the location where the particle acceleration initiates $z_{\rm{diss}}$, the plasma beta parameter $\beta$, the parameters for the disc, namely the innermost radius $R_{\rm{in,disc}}$ and the mass accretion rate in Eddington units ($\dot{m} = \dot{M}{\rm{c}}^2/L_{\rm{Edd}}$),  and the parameters of the corona, namely the temperature $T_{\rm{cor}}$, the normalisation radius $R_{\rm{cor}}$ and the optical depth $\tau_{\rm{cor}}$.
 
 %%%%%%%%%%%%%%%%%%%%%%%%%%%%%%%%
%%%% FIXED PARAMETERS TABLE %%%%
%%%%%%%%%%%%%%%%%%%%%%%%%%%%%%%%
\begin{table}
	\begin{center}
		\setlength{\tabcolsep}{3pt} % Default value: 6pt
		\renewcommand{\arraystretch}{1.4} % Default value: 1
		\begin{tabular}{lrl}\hline
			parameter & value & description \\
			\hline
			$M_{\rm{BH}}\, \left(\rm{M_{\odot}}\right)$& 21.4      & mass of the black hole$^{\dagger}$\\
			$\theta_{\rm{incl}}$ 		& 27.5$^\circ$        & viewing angle$^{\dagger}$\\
		$D\, \rm{\left(kpc\right)}$ 				& 2.22              & distance of the source$^{\dagger}$\\
			$N_H\, (10^{22}$\,cm$^{-2})$&0.6& number column density\\
			$h=z_0/R_0$ 			& 2                 & initial jet height to radius ratio \\
			$z_{\rm{max}}\, \left(r_{\rm{g}}\right)$ 		& $10^8$            & maximum jet height$^{\ast}$\\
			$T_{\rm{\star}}\, \left(K\right)$ 		& $3.08\times 10^4$  & temperature of the companion star$^{\dagger}$ \\
			$L_{\rm{\star}}\, \rm{\left(erg\,s^{-1}\right)}$ 	& $1.57\times 10^{39}$ & luminosity of the companion star$^{\dagger}$\\
			$a_{\star}\, \rm{\left(cm\right)}$ 		& $3.7\times 10^{12}$& orbital separation distance$^{\dagger}$\\
			$\dot{M}_{\star}\, \left(\rm{M_{\odot}}\,\rm{yr^{-1}}\right)$ & $2.6\times 10^{-6}$     & mass loss rate of the companion star$^{\ddagger}$\\
			$v_{\rm{wind}}\, \rm{\left(cm\,s^{-1}\right)}$    & $2.4\times 10^8$    & velocity of the stellar wind$^{\ddagger}$ \\ 
%			$R_{\rm{out,disc}}\, \left(r_{\rm{g}}\right)$ 	& $10^{5}$          & disc outermost radius\\
			\hline
		\end{tabular} 
		\caption{The fixed parameters of our models. $^{\dagger}$Miller-Jones et al. (subm), $^{\ast}$\protect\cite{tetarenko2019radio}, %$^{\star}$\protect\cite{orosz2011mass},
		$^{\ddagger}$similar to \protect\cite{Grinberg2015variability}.} 
		\label{table:common parameters}
	\end{center}
\end{table}

We present here the results of the best fits of our models. We choose one lepto-hadronic and one purely leptonic model to reproduce the MeV X-rays as jet synchrotron radiation, so as to explain the high degree of linear polarization \citep{2011Sci...332..438L,jourdain2012separation,0004-637X-807-1-17,Cangemi2020longterm}. We achieve this by assuming that the non-thermal electrons accelerate in a hard power-law. We find that an index of $p=1.7$ provides sufficient results. We show two more models for comparison. One purely leptonic and one lepto-hadronic, with softer power-laws of $p=2.2$. With such an assumption we fail to reproduce the MeV polarization as we show below.

\subsection{Plasma quantities}
The four different models presented here lead to different jet dynamical quantities, as we show in Table~\ref{table: parameters for models}. The jet base radius varies between 2--27$\,r_{\rm{g}}$ and the region where the energy dissipates into particle acceleration varies between 15--125$\,r_{\rm{g}}$. The two models with a hard injected particle distribution require a small value of plasma  $\beta$ compared to the softer models.

The best-fitting values for the injected power $L_{\rm{jet}}$ for the models with the hard power law ($p=1.7$), are comparable. Based on the jet-base radius $R_0$ and the plasma $\beta$, we calculate the strength of the magnetic field along the jet. For all our models, we find relatively high magnetic field strengths at the jet base on the order of $10^6\,$G. 

\begin{figure*}
\begin{center}
	%\centering
	%\subfloat{
	\includegraphics[width=1.1\columnwidth]{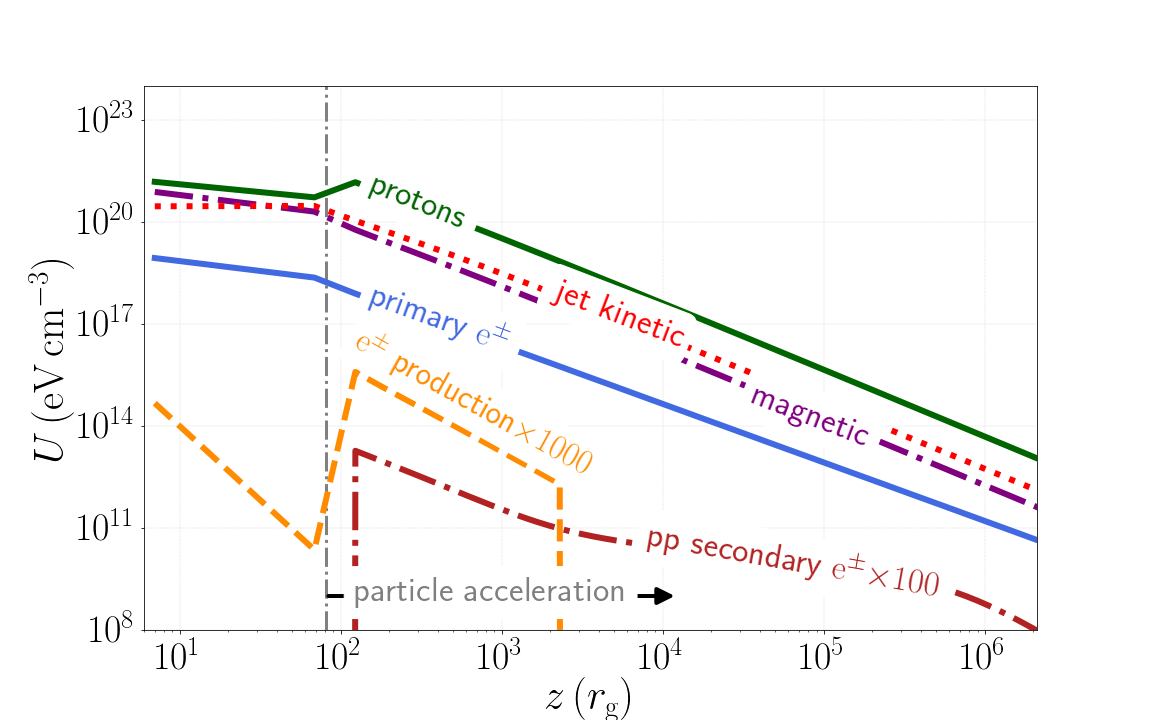}%}
	%\subfloat{
	%\\
	\includegraphics[width=1.1\columnwidth]{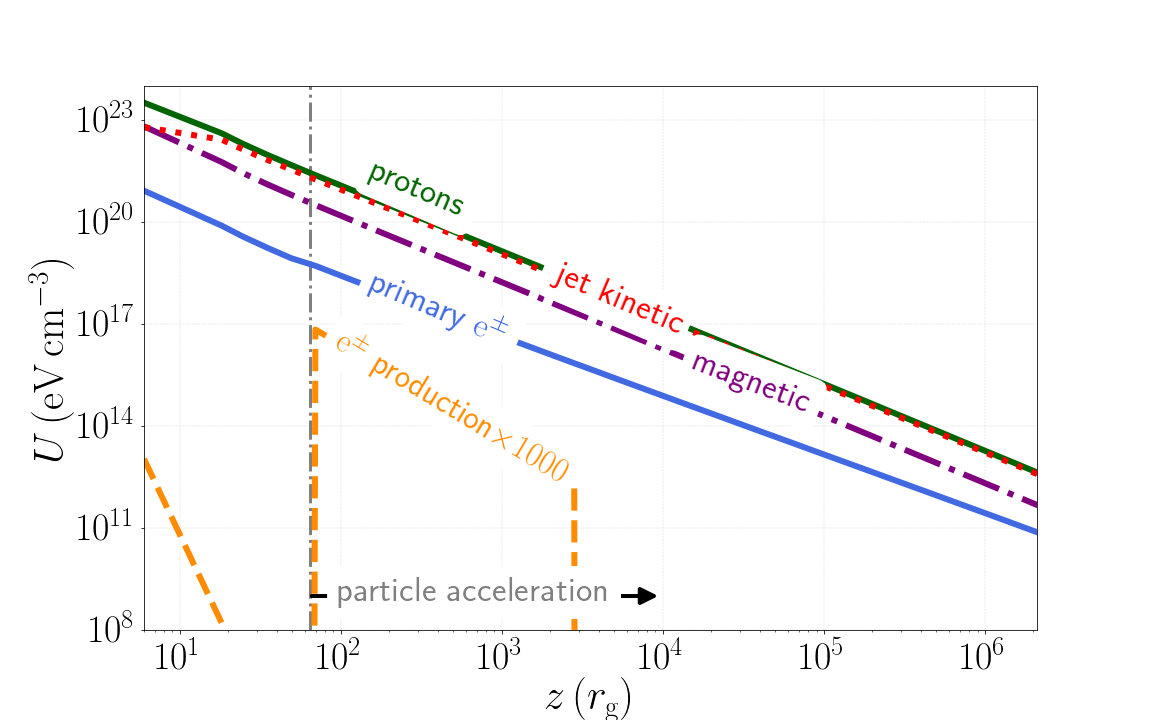}%}
	\caption{Contributions to the total energy density as a function of the distance along the jet for the model with a power-law index $p=1.7$, for the the lepto-hadronic case ({\bf left}) and the purely leptonic case ({\bf right}). The particle acceleration initiates at the vertical dot-dashed grey line. The jump in the proton energy density on the left plot is due to proton acceleration. We do not assume extraction of energy from other components to accelerate the particles. The proton and the jet kinetic energy density of the right plot coincide because no proton acceleration is taken into account. We stop to calculate the pair production after some distance because it has insignificant contribution. }
	\label{fig:energy density along jet}
\end{center}
\end{figure*}

In Fig. \ref{fig:energy density along jet} we plot the energy density of various quantities along the jet axis for models the two models with $p=1.7$. In particular, our fits are driven towards particle-dominated jets with the energy density of the protons dominating along the jet. Moreover, the energy density of the magnetic field is higher than the energy density of the (primary) electrons. We also show the energy density of the secondary pairs due to photon annihilation. We see that this process has its peak but still insignificant contribution in jet segments of high compactness, i.e. high photon number density at the jet base and in the particle acceleration region. The number density of the target photons drops significantly after the jet base, which suppresses the pair production. At the particle acceleration region the compactness increases due to the non-thermal synchrotron and SSC photons. For the case of the lepto-hadronic model, we also show the energy density of secondary electrons from pp interactions, even though their energy density is more than five orders of magnitude lower than the rest. 

%%%%%%%%%%%%%%%%%%%%%%%%%%%%%%%
%%%% FREE PARAMETERS TABLE %%%%
%%%%%%%%%%%%%%%%%%%%%%%%%%%%%%%
\begin{table*}
	\begin{center}
		\setlength{\tabcolsep}{10pt} % Default value: 6pt
		\renewcommand{\arraystretch}{1.5} % Default value: 1
		\begin{tabular}[b]{lrrrrl}\hline
			parameter & \multicolumn{2}{c}{lepto-hadronic models} &\multicolumn{2}{c}{leptonic models}  &description \\
			\hline
			$p$    & 		1.7&2.2& 1.7&2.2&particle power-law index\\
			%$f_{\rm{sc}}$ 	&	$0.01$&0.1& 0.05&0.1&particle acceleration parameter\\					
			\hline	
			\multirow{2}{*}
			{$L_{\rm{jet}} \begin{array}{c}
			      \left(10^{-4}\,L_{\rm{Edd}}\right)  \\ 
			      \rm{\left(erg\,s^{-1}\right)}
			\end{array}$
		} &$9_{-5}^{+26}$  & $105_{-6}^{+360}$ &$7.1_{-2.5}^{+3.4}$ & $2.0_{-1.9}^{+2.1}$ & \multirow{2}{*}{jet base injected power} \\
			&$ 2.4_{-1.3}^{+7.0}\times 10^{36}$& $ 28.3_{-1.6}^{+97.1}\times 10^{36}$ & $ 19.1_{-6.7}^{+9.2}\times 10^{35}$ & $ 5.4_{-5.1}^{+5.7}\times 10^{35}$&\\
			$R_0 \left(r_{\rm{g}}\right)$            & $ 27_{-25}^{+29}$    & $ 2_{}^{+21}$ & $ 3.1_{-1.1}^{+12.5}$&$ 3.1_{-1.1}^{+4.1}$ &jet base radius\\
			$z_{\rm{diss}}\,\left(r_{\rm{g}}\right)$ 		        & $ 81_{-15}^{+114}$& $ 15_{-4}^{+496}$      & $64 _{-14}^{+57}$ & $ 125_{-45}^{+475}$ & particle acceleration region\\
			$T_{\rm{e}}\,(\rm{keV})$ &$1762_{-1267}^{+3375}$ & $711_{-703}^{+726}$ &$756_{-726}^{+1530}$ & $1114_{-1096}^{+1401}$& jet base thermal electron temperature\\
			$\beta$         & $ 0.011_{-0.001}^{+0.529}$ & $ 0.18_{-0.02}^{+0.89}$      & $0.013 _{-0.011}^{+0.332}$ & $ 0.95_{-0.84}^{+0.97}$& plasma beta\\
			$\dot{m}\,(10^{-3})$ 	& $ 2.5_{-2.4}^{+15.9}$ &$ 1.22_{-1.20}^{+1.25}$              & $ 1.4_{-1.3}^{+1.5}$& $ 1.1_{-1.0}^{+1.2}$& mass accretion rate\\
			$R_{\rm{in,disc}}\, \left(r_{\rm{g}}\right)$    & $ 19_{-17}^{+11}$ &$ 3.1_{-1.1}^{+3.3}$&$ 6.3_{-4.3}^{+6.4}$&$ 3.1_{-1.1}^{+3.3}$&disc innermost radius\\
			$T_{\rm{cor}}\, \rm{\left(keV\right)}$ 		    & $ 90_{-10}^{+69}$ & $ 89_{-9}^{+16}$       & $ 105_{-14}^{+51}$ & $ 81_{-1}^{+82}$&corona temperature\\
			$R_{\rm{cor}}\, \left(r_{\rm{g}}\right)$ 		    & $ 59_{-19}^{+60}$ & $10_{-6}^{+38}$  & $ 20_{-9}^{+41}$ & $ 10_{-6}^{+9}$&corona normalisation radius\\
			$\tau_{\rm{cor}}$ 			        & $ 0.58_{-0.47}^{+0.32}$ &$ 0.62_{-0.60}^{+0.63}$    &$ 0.49_{-0.39}^{+0.50}$ &$ 0.61_{-0.59}^{+6.30}$ &corona optical depth\\
			\hline \hline
			$B_0\,\left(\rm{G}\right)$ 			&$1.8\times 10^6$   & $1.6\times 10^6$    & $1.6\times 10^{6}$ &$6.4\times 10^5$&magnetic field at jet base\\
			$B\,\left(\rm{G}\right)@\, z_{\rm{diss}}$ 	&$9.1\times 10^4$  &$1.0\times 10^6$     & $1.6\times 10^{5}$ & $3.2\times 10^4$ &magnetic field at particle acceleration region \\
			$L_p\,\rm{\left(erg\,s^{-1}\right)}$ 		&$4.3\times 10^{38}$&$5.1\times 10^{38}$    &  -& - &accelerated proton power\\
			$L_e\,\rm{\left(erg\,s^{-1}\right)}$ 		&$1.3 \times 10^{36}$ & $1.9 \times 10^{36}$   & $2.1 \times 10^{36}$ & $3.3 \times 10^{36}$&accelerated electron power\\
			$E_{p,\rm{max}}\,\rm{\left(eV\right)}$ 	&$2.7\times 10^{15}$& $1.8\times 10^{15}$    & -&-&proton maximum energy\\
			$E_{e,\rm{max}}\,\rm{\left(eV\right)}$ 	&$6.1\times 10^{10}$&$2.2\times 10^{10}$    &   $3.3\times 10^{10}$ & $8.8\times 10^{10}$&(primary) electron maximum energy\\
			\hline
			$\chi^2/DoF$&  9597.8/2439& 2451.9/2439 & 4064.7/2439 &2237.1/2439  & $\chi^2/$degrees of freedom \\
		\end{tabular} 
		\caption{The free parameters of the four models discussed in this paper that differ in the power-law index $p$ of the accelerated particles. Before the double line, we show the fitted parameters and their uncertainties. Below, we show the evaluated quantities of the magnetic field, the total luminosity of the accelerated proton/electron population and the maximum energy of the protons/electrons at the particle acceleration region.}
		\label{table: parameters for models}
	\end{center}
\end{table*}

%%%%%%%%%%%%%%%%%%%%%%%%%%%%%%%%%%%%%%%%
%%%% ENERGY DENSITIES ALONG THE JET %%%%
%%%%%%%%%%%%%%%%%%%%%%%%%%%%%%%%%%%%%%%%
\begin{comment}
\begin{figure*}
	\centering
	\subfloat{\includegraphics[width=1.1\columnwidth]{figs/HH/U_z.pdf}}
	\subfloat{\includegraphics[width=1.1\columnwidth]{figs/LH/U_z.pdf}}
	\caption{Energy density along the jet...}
	\label{fig:energy density along jet}
\end{figure*}
\end{comment}

%%%%%%%%%%%%%%%%%%%%%%%%%%%%%%%%%%
%%%% HADRONIC SPECTRA FIGURES %%%%
%%%%%%%%%%%%%%%%%%%%%%%%%%%%%%%%%%
\begin{figure*}
\begin{center}
	\includegraphics[width=1.7\columnwidth]{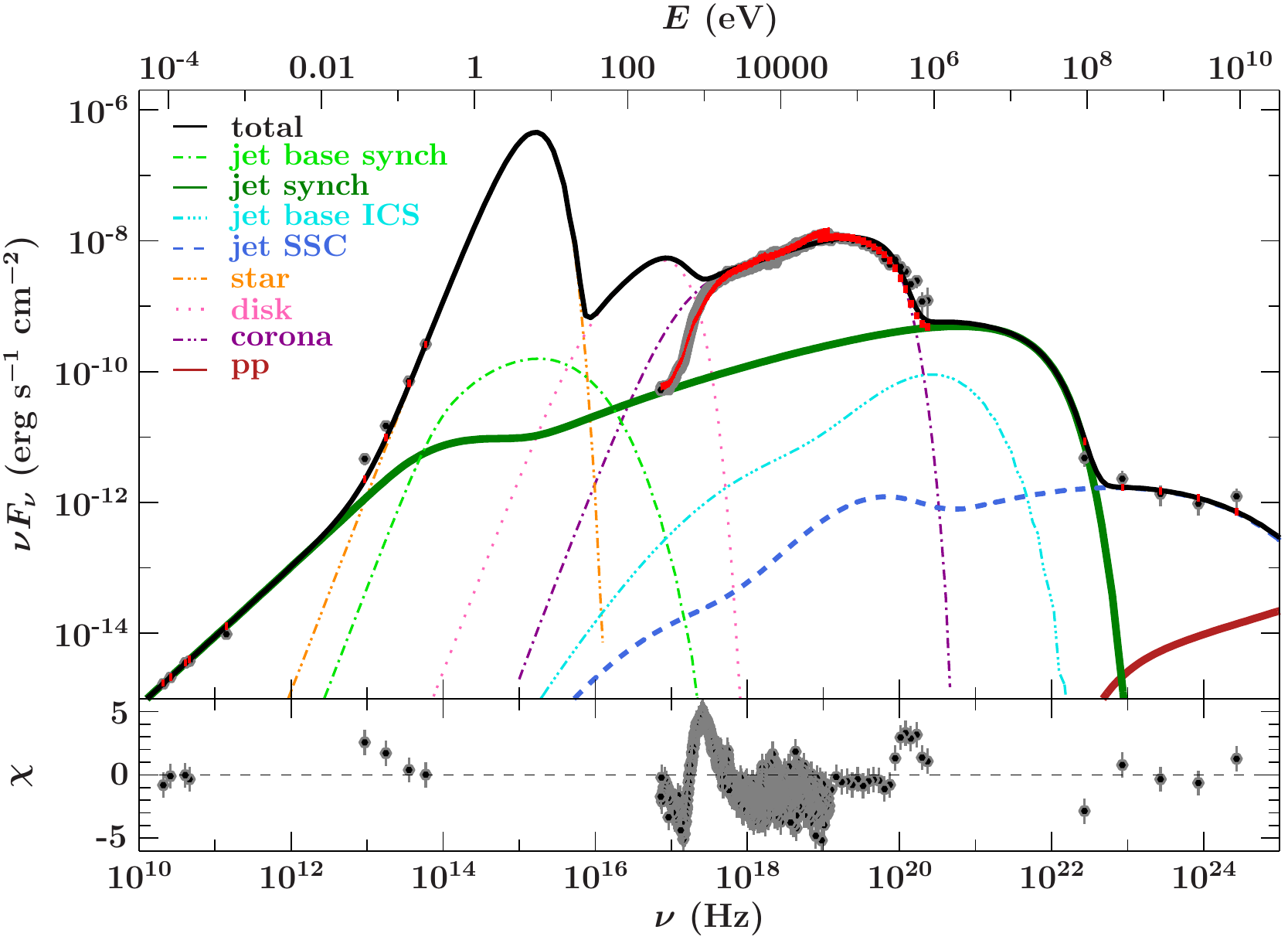}\\
	\includegraphics[width=1.7\columnwidth]{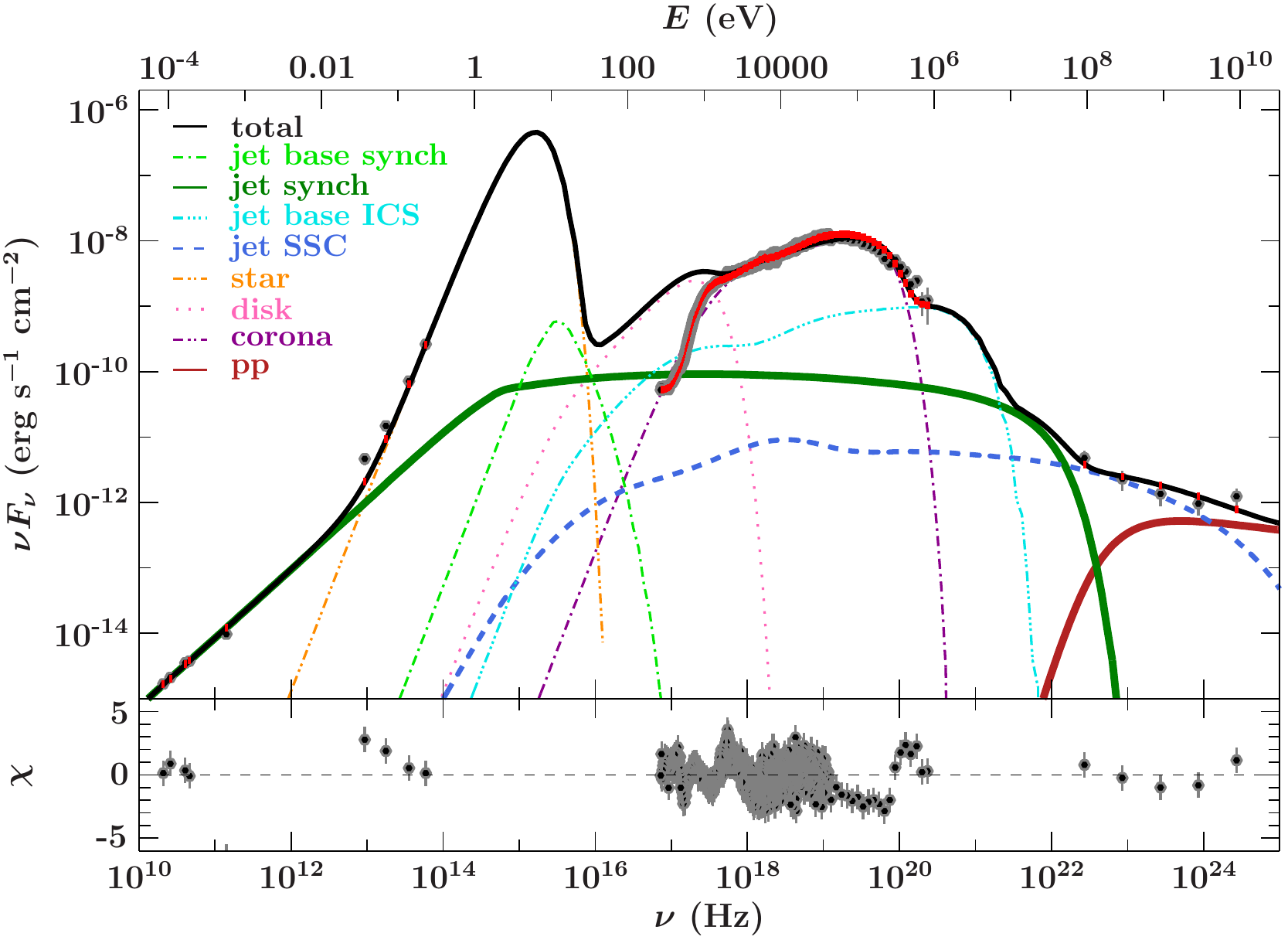}
	\caption{The best-fit multiwavelength spectrum of Cyg~X--1 for the two lepto-hadronic scenarios with $p=1.7$ ({\bf top}) and $p=2.2$ ({\bf bottom}) and their $\chi$ residuals. The solid black line shows the total unabsorbed spectrum. The absorbed spectrum that we fitted to the data in detector space is shown as solid red line. We also show some individual unabsorbed model components, i.e. the broad-band radio-to-$\g$-ray synchrotron spectrum from primary electrons (thick solid green line), the ICS spectrum ranging from eV to GeV (dashed dark blue line), the pp spectral component arising from the neutral pion decay (solid red line), disc photons upscattered in the thermal corona (dotted-dashed purple line), the black-body component emitted by the companion star (double-dotted-dashed orange line), and the multi-temperature thermal spectrum arising from the accretion disk (dotted magenta line). The dotted-dashed light green line shows the synchrotron radiation from thermal electrons and the triple-dotted-dashed light blue line shows the ICS from regions before the particle acceleration region. In the case where $p=1.7$ the jet-synchrotron dominates in the MeV band explaining the high degree of reported linear polarisation. In the soft case of $p=2.2$, the fit does not explain the reported polarisation.}
    \label{fig:spectrum with hadronic processes}
\end{center}    
\end{figure*}

%%%%%%%%%%%%%%%%%%%%%%%%%%%%%%%%%%
%%%% LEPTONIC SPECTRA FIGURES %%%%
%%%%%%%%%%%%%%%%%%%%%%%%%%%%%%%%%%
\begin{figure*}
\begin{center}
	\includegraphics[width=1.7\columnwidth]{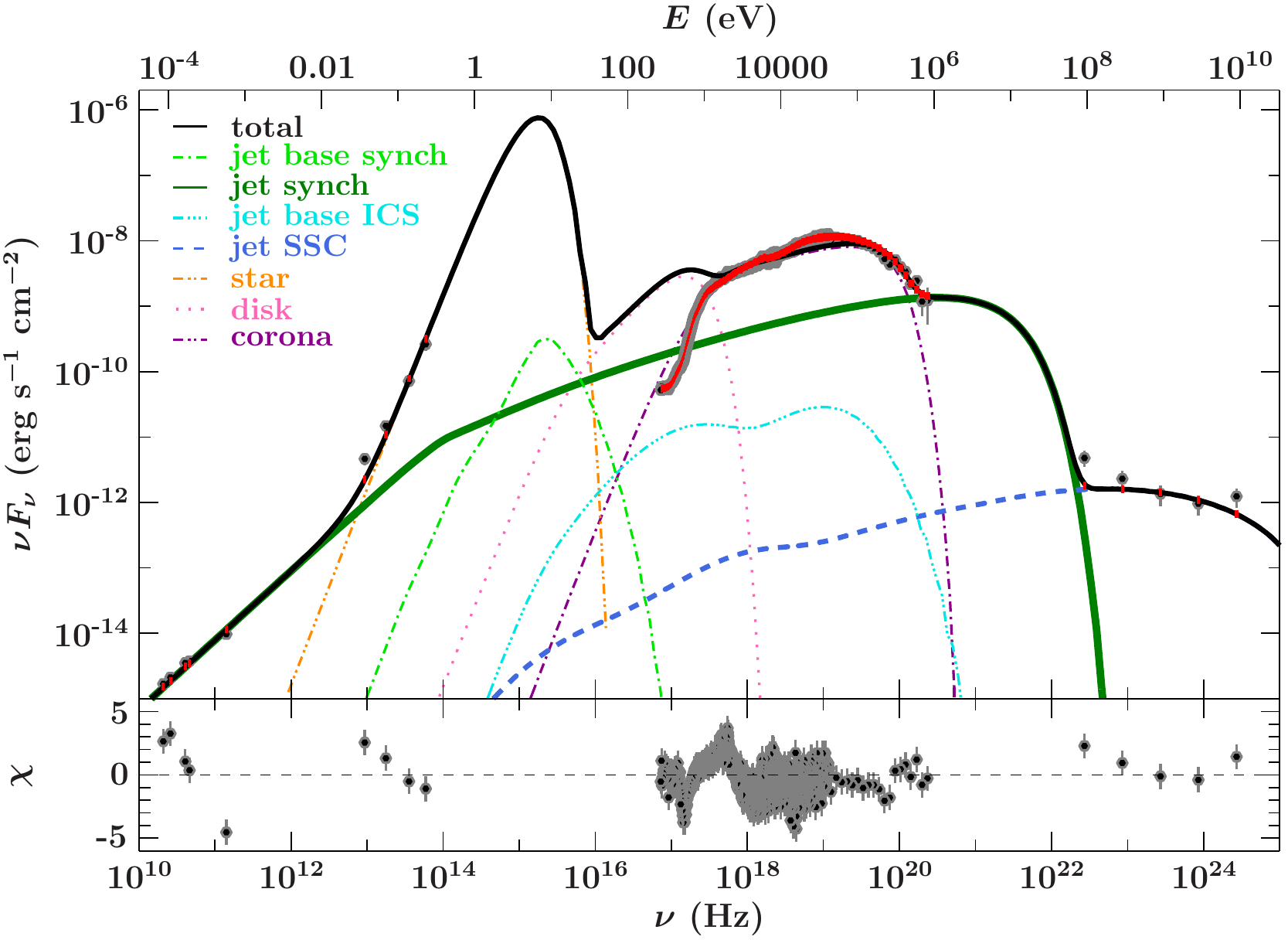}	
	\includegraphics[width=1.7\columnwidth]{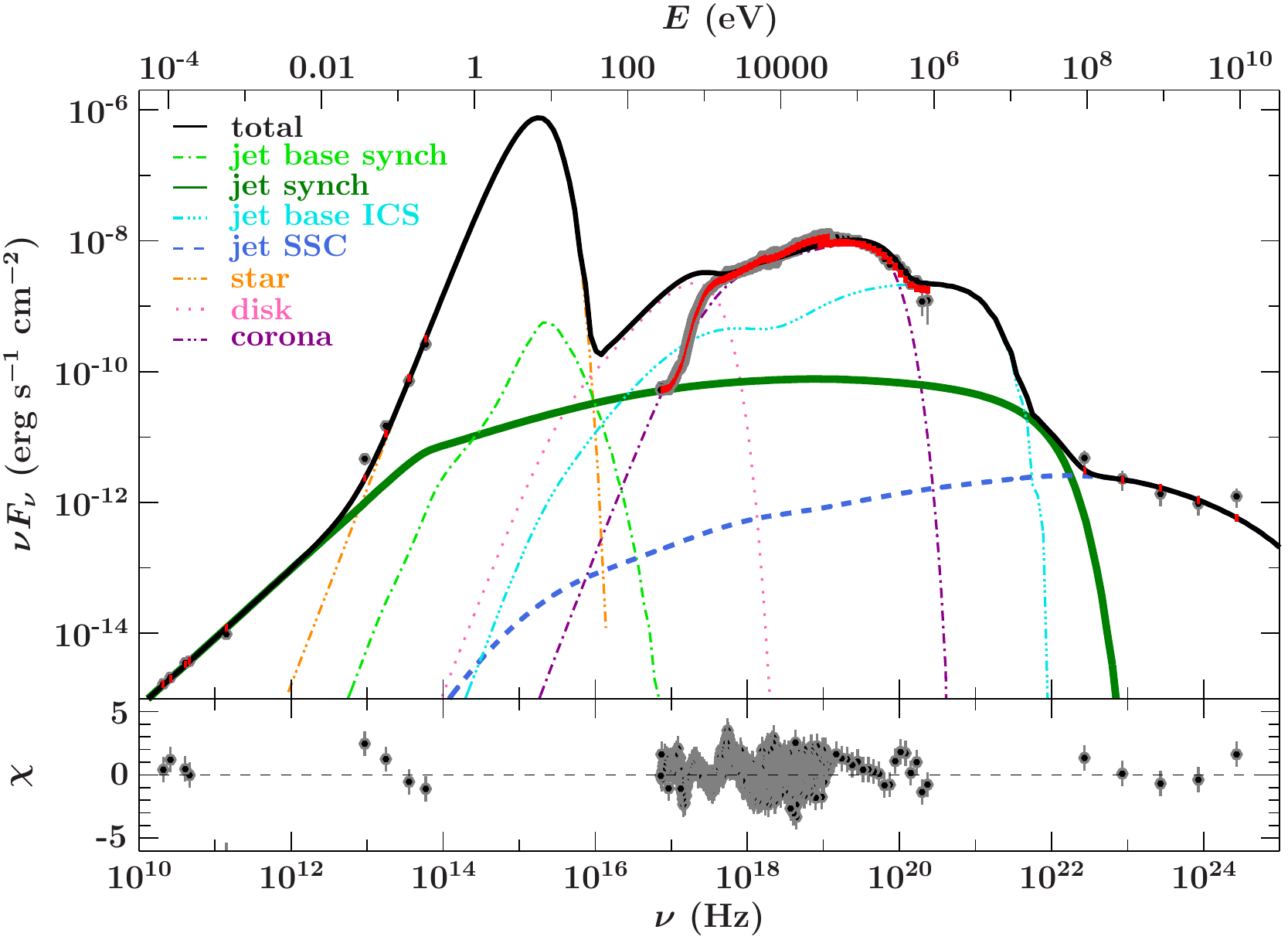}	
    \caption{Similar to Fig. \ref{fig:spectrum with hadronic processes} but for the two leptonic scenarios with $p=1.7$ ({\bf top}) and $p=2.2$ ({\bf bottom}).}
    \label{fig:spectrum without hadronic processes}
\end{center}    
\end{figure*}

%%%%%%%%%%%%%%%%%%%%%%%%%%%%%%%%%%%%
%%%% GeV/TeV ZOOMED-IN SPECTRUM %%%%
%%%%%%%%%%%%%%%%%%%%%%%%%%%%%%%%%%%%
\begin{figure*}
\begin{center}
	\includegraphics[width=2.1\columnwidth]{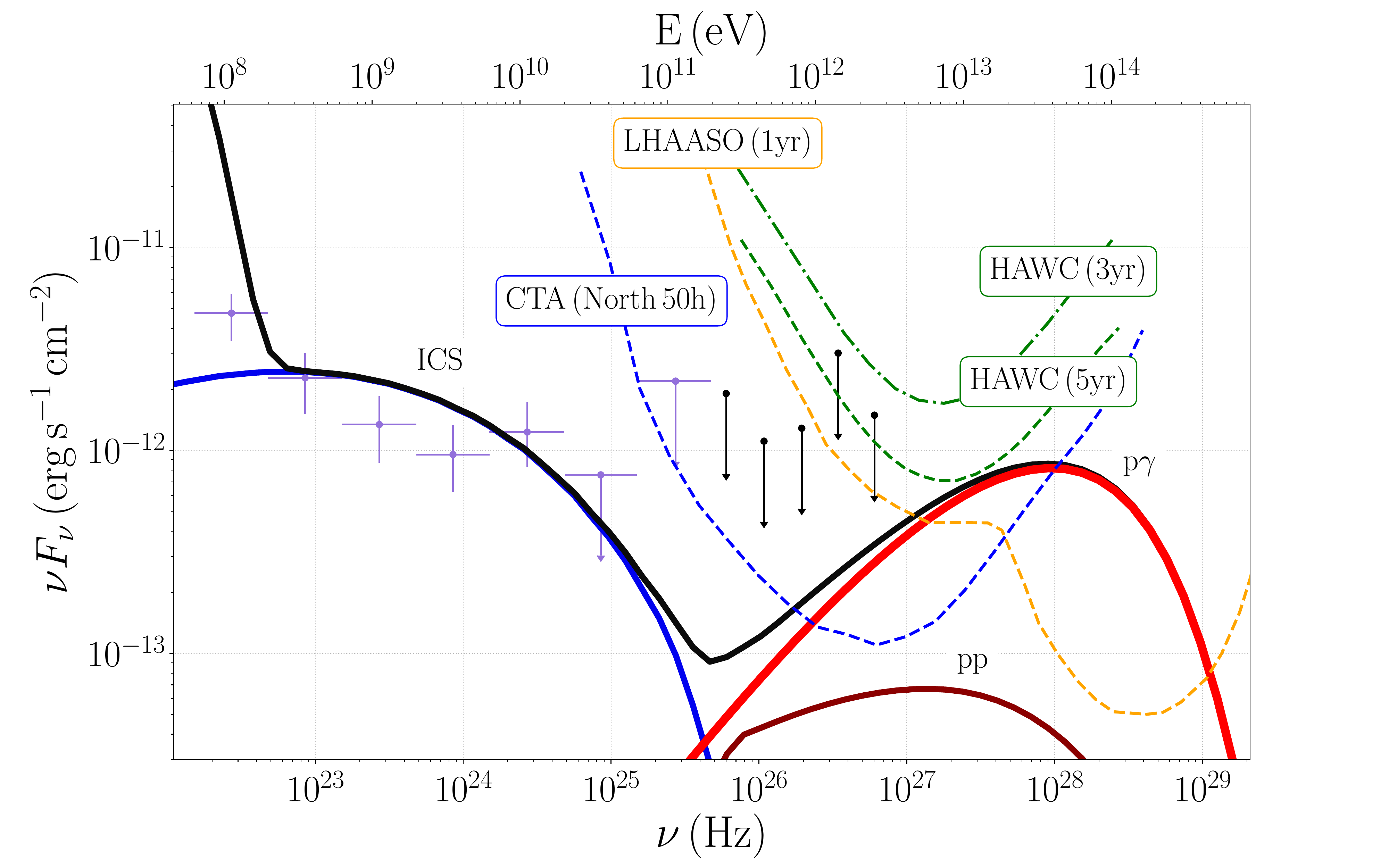}
    \caption{The GeV-to-TeV regime of the multiwavelength spectrum of Cyg~X--1 for the lepto-hadronic scenario with $p=1.7$. The black line shows the total spectrum. The ICS (solid dark blue) explains the \textit{Fermi}/LAT (purple) data points in the GeV band. The neutral pion decay from p$\g$ (thick light red) dominates the pp (dark red) and peaks in the TeV regime. Such emission will be detectable by future generation facilities, such as the CTA (dashed blue, adopted from \href{www.cta-observatory.org}{www.cta-observatory.org}), and LHAASO (dashed orange, adopted from \protect\citealt{bai2019large}). We also plot for comparison the upper limits of MAGIC (black upper limits) from \protect\cite{ahnen2017search}, and the 3 (dot-dashed green) and 5-year (dashed green) sensitivity of HAWC adopted  from \protect\cite{HAWC2013sensitivity}. }.
    \label{fig:spectrum with hadronic processes zoomed}
\end{center}    
\end{figure*}

\subsection{Best fits to the multiwavelength spectrum}\label{section: best fits}
The combined data of Cyg~X--1 presented in Section~ \ref{section: observational data} result in a broad-band spectrum covering almost 15 orders of magnitude in photon frequency. We are able to reasonably fit all wavebands simultaneously with our model. Figures~\ref{fig:spectrum with hadronic processes} and \ref{fig:spectrum without hadronic processes} show all four different model scenarios. The residuals are not always negligible, especially for the X-ray spectrum between $10^{17}$ and $10^{19}\,$Hz. This is a natural consequence of our broad-band fit. The superb data coverage of the X-rays suggests a number of specific spectral features, e.g., due to relativistic reflection off the inner accretion disc, which our over-simplified model for the corona is not able to describe in detail. Such an in-depth treatment of all X-ray features is outside the scope of this work (see e.g., \citealt{Tomsick_2013,duro2016revealing,Parker_2015,basak_2017}).

We also take into account synchrotron-self absorption in the radio band and photoabsorption of X-ray photons with the column density $N_H=0.6\times 10^{22}\,$cm$^{-2}$ \citep{Grinberg2015variability}. The wind of the companion star could in principle attenuate the radio band even at inferior conjunction (when the companion star is behind the jet on the line of sight) examined here. Nevertheless, the 20$\,$GHz radio emission originates from a region much further out in the jets than 10 times the separation of the system so this attenuation should be insignificant \citep[see e.g.,][]{szostek2007jet}. 

\subsection{GeV-TeV spectrum}\label{section: GeV-TeV spectrum}
The lepto-hadronic model with $p=1.7$ is the only one that predicts significant TeV emission. In Fig.~\ref{fig:spectrum with hadronic processes zoomed} we plot the GeV to $\sim$ PeV regime of its multiwavelength spectrum. For a comparison, we add the upper limits of the Major Atmospheric Gamma Imaging Cherenkov Telescopes - MAGIC \citep{ahnen2017search}, the 3 and 5-year sensitivity of the High-Altitude Water Cherenkov Observatory - HAWC \citep{HAWC2013sensitivity}, and the predicted sensitivity of the Cherenkov Telescope Array - CTA (from \href{www.cta-observatory.org}{www.cta-observatory.org}) and of the Large High Altitude Air Shower Observatory - LHAASO \citep{bai2019large}.

In the GeV range, we did not take into account photon annihilation due to the stellar photon field because the data we consider here are taken while the source was in the inferior conjunction. Further GeV observations will help to better understand the orbital modulation of Cyg X-1 as well in this domain. 

Our evaluated spectrum above $0.1\,$TeV ($10^{25}\,$Hz) is dominated by the $\g$-rays produced via neutral pion decay from the hadronic collisions. The dominant process at the highest photon energies is the p$\g$ interaction, between accelerated jet protons and the synchrotron MeV photons. The number density of other target photon fields is negligible compared to this MeV band in the jet rest frame. The flux levels predicted by our model are overall higher than the sensitivity limits of next-generation $\g$-ray telescopes. HAWC, LHAASO, and CTA will therefore be key for breaking further degeneracies within our model, and constraining important processes such as the p$\g$ interactions in astrophysical jets.

For our discussion of the highest energies, we only consider the hard lepto-hadronic model ($p=1.7$), as the soft model ($p=2.2$) cannot explain the MeV polarisation. Neither leptonic model can produce any TeV emission via ICS, because the electron scattering with GeV photons occurs deep in the Klein-Nishina regime. Thus, no further order scatters can occur inside the jets that would produce significant TeV radiation. A solid TeV detection would therefore rule out the leptonic models.

\section{Discussion}\label{section: discussion}

A key open issue regarding Cyg~X--1 is the polarised 0.4--2\,MeV tail detected by \textit{INTEGRAL} while the source is in the hard state \citep{2011Sci...332..438L,jourdain2012separation,0004-637X-807-1-17,Cangemi2020longterm}. The above studies all independently conclude that the linear polarisation degree of the MeV emission is of the order of 50--70 per cent. While there is an overall agreement on the degree of polarisation, \integral\ does not have the spatial resolution to resolve the source, thus the integrated polarisation angle over the entire system does not provide constraining information on the detailed magnetic field geometry of the source.

Such high degree of polarisation, requires a structured and well-ordered magnetic field. High-resolution numerical simulations suggest that the wind of the accretion disc, which is associated to the corona, is very turbulent and could not explain such structured magnetic field \citep{chatterjee2019accelerating,liska2017,liska2020}. Hence, jet-synchrotron is more likely to explain the MeV polarisation.

In this work, we take advantage of the new and unprecedented (in broadband simultaneity) CHOCBOX multi-wavelength data set to revisit the question of leptonic vs. hadronic processes, using a more sophisticated multi-zone approach.  In particular we explore the consequences of taking the MeV polarisation as a `hard' constraint, and the consequences for potential TeV $\gamma$-ray emission. We find that the only way to produce sufficient synchrotron flux to fit the MeV data is by assuming a hard power-law distribution of accelerated electrons with $p=1.7$. If we assume a soft power-law with $p=2.2$ we fail to match this constraint.

These two different power-law indices of 1.7 and 2.2 are typically associated with different particle acceleration mechanisms. The hard particle spectrum ($p=1.7$) suggests second-order Fermi acceleration \citep[e.g.,][]{rieger2007fermi} or magnetic reconnection (e.g., \citealt{biskamp1996magnetic,Sironi_2014,petropoulou2018steady} or \citealt{khiali2015magnetic} for the case of Cyg~X--1 specifically). The softer injection value of $p=2.2$ is more suggestive of non-linear diffusive shock acceleration  \citep[e.g.,][]{drury1983introduction,malkov2001nonlinear,Caprioli2012nldsa}, but we show that the high degree of MeV polarisation cannot be attained. We find that the best fits to the data require a more efficient acceleration mechanism to be the dominant source of non-thermal particles. We note however that when we define the acceleration timescale to derive the maximum energy of the particles (see Equations~\ref{proton max energy} and \ref{electron max energy}), we use a simplified expression that is commonly used to describe first-order Fermi acceleration. In future work, we will include energy dependence to the acceleration timescale to explore in detail the different acceleration mechanisms.

Taking as a constraint the explanation of both the observed MeV spectrum and the GeV $\g$-rays, we require a generally high particle acceleration efficiency $f_{\rm{sc}}$. For the models with a soft particle spectrum, we require a higher efficiency (0.1) as opposed to the models with the hard particle spectrum, where an acceleration efficiency of 0.01 is sufficient. This parameter also drives the maximum achievable energy of the particles. We find a maximum electron energy of 10--100\,GeV (see Table~\ref{table: parameters for models}) and proton energy of $\sim 10^{15}$\,eV. The high particle energies we find for both electrons and protons translate to a required high total power in particles, i.e. $\sim 10^{36}\,\rm{erg\,s}^{-1}$ for electrons and $\sim 10^{39}\,\rm{erg\,s}^{-1}$ for protons. 

Independent measurements of the total kinetic jet power are useful to benchmark our fitted values for the total injected energy. One can estimate the jet power from the bubble-like structure located 5\,pc from Cyg~X--1 caused by the apparent interaction between the jet and the ISM. The mechanical power required to inflate such bubble has been calculated to be of the order of $10^{37}\,\rm{erg\,s^{-1}}$ \citep{gallo2005dark}.  It is, however, still debated whether the jet is solely evacuating this bubble, or whether other feedback channels, such as the companion star's stellar wind, play a role. In that case, the jet power estimated by \cite{gallo2005dark} would have to be considered as an upper limit \citep{sell2015shell}. This estimate would lead to the exclusion of the lepto-hadronic model because of its exceeding jet power, while the purely leptonic model requires merely 10 percent of the estimated power. This large discrepancy (up to 3 order of magnitude) driven by the inclusion/exclusion of hadronic processes is a well-known issue in the field \cite[e.g.,][]{bosch2008magnetic,zdziarski2012mev,malyshev2013high,zdziarski2014jet,Zhang_2014,pepe2015lepto,zdziarski2017high,Beloborodov_2017,fernandez2017gamma}.

Most hadronic models show jet powers close to Eddington limit either for Galactic or extragalactic sources \citep{boettcher2013leptohadronic,zdziarski2015hadronic}. However, there are a few possible ways of extracting further power from the system to the particles without violating other constraints. One possibility is a much more efficient dissipation of either magnetic or kinetic energy via particle acceleration, i.e. greater than 10 per cent. Another, perhaps more likely scenario is the one where the jets are launched by a magnetically dominated (MAD) accretion flow and a spinning black hole. In such systems, the jet can benefit from an efficient extraction of power both from the accretion disc and the black hole rotation \citep{blandford1977extraction,narayan2003mad,Tchekhovskoy2011efficient}. Alternatively, the total proton power can be reduced. One possibility is that the jets are predominantly leptonic up to when the bulk flow is accelerated to maximum velocity. The majority of protons are then mass-loaded further away from the launching point either by the wind of the accretion disc or of the companion star \citep{chatterjee2019accelerating,perucho2020mixing}. To calculate the total proton power in this work, we sum the proton power per segment along the jet. If we assume that protons accelerate only within a small part of the jet, then the total power could be significantly reduced  \citep{pepe2015lepto,khiali2015magnetic,abeysekara2018very}. Such assumptions would however only increase the free parameters of our model. Therefore, we decided to restrict ourselves to 'standard' assumptions for fitting the data, and to ease comparison with prior approaches.

\subsection{Comparison with previous works}

%%%%%%%%%%%%%%%%%%%%%%%%%%
%%%% COMPARISON TABLE %%%%
%%%%%%%%%%%%%%%%%%%%%%%%%%
\begin{table*}
	\begin{center}
		\setlength{\tabcolsep}{6pt} % Default value: 6pt
		\renewcommand{\arraystretch}{1.5} % Default value: 1
		\begin{tabular}{>{\arraybackslash}m{0.44\columnwidth} >{\centering\arraybackslash}m{0.15\columnwidth}  >{\centering\arraybackslash}m{0.18\columnwidth}
		>{\centering\arraybackslash}m{0.18\columnwidth}
		>{\centering\arraybackslash}m{0.18\columnwidth} >{\centering\arraybackslash}m{0.18\columnwidth}| >{\centering\arraybackslash}m{0.18\columnwidth} >{\centering\arraybackslash}m{0.18\columnwidth}}\hline
        &
			\multicolumn{5}{c}{other works}
			& \multicolumn{2}{c}{this work}\\ 
			features \textbackslash  model& R14 & Z14 & K15 & P15 & Z17 & hadronic & leptonic\\
			\hline \vspace{2mm}
power-law index$^{\dagger}$&2.2& $1.4/2.5$& 1.8& $2.0/2.4$& 2.2 & $1.7/2.2$ & $1.7/2.2$\\
corona presence & \cmark& \cmark& \xmark & \cmark & \cmark& \cmark& \cmark\\
hadronic processes & \cmark & \xmark &\cmark & \cmark & \xmark& \cmark & \xmark\\
simultaneous data & \xmark & \xmark &\xmark& \xmark &\xmark &\cmark&\cmark \\
statistical modelling / MCMC &  \xmark &  \xmark &\xmark& \xmark &\xmark &\cmark&\cmark \\
MeV X-rays origin& cor-SYN& SYN/COM & SYN & SYN/COM & COM &SYN/COM &SYN/COM\\
explain MeV polarisation & \cmark & \cmark/\xmark & \xmark & \cmark/\xmark & \xmark & \cmark/\xmark &\cmark/\xmark\\
CTA @ TeV prediction &\xmark&\xmark&\cmark&\cmark/\xmark&\xmark&\cmark&\xmark\\
LHAASO @ 100\,TeV prediction& \xmark& \xmark&(\cmark)&(\cmark)/\xmark&\xmark&\cmark &\xmark\\
			\hline
		\end{tabular} 
		\caption{Comparison between our results and previous works on reproducing the (multiwavelength) spectrum of Cyg~X--1. When two models are discussed in a specific work, we separate them with a slash. cor-SYN stands for synchrotron radiation from a non-thermal corona, SYN for jet (primary) synchrotron radiation and (\cmark) stands for a marginal detection. References included are: \citealt{romero2014MeVtail} (R14), 		\citealt{zdziarski2014jet} (Z14), \citealt{khiali2015magnetic} (K15), \citealt{pepe2015lepto} (P15) and \citealt{zdziarski2017high} (Z17).\\
$^{\dagger}$accelerated particle power-law
index $p$: $N(E)\propto E^{-p}$.
}
		\label{table:comparison between models}
	\end{center}
\end{table*}

In Table~\ref{table:comparison between models} we present a schematic comparison between the main features of our new model and of a sample of similar works used to explain the multiwavelength spectrum of Cyg~X--1. The models that we consider here are the following: \citealt{romero2014MeVtail} (R14),
\citealt{zdziarski2014jet} (Z14), \citealt{khiali2015magnetic} (K15), \citealt{pepe2015lepto} (P15), and \citealt{zdziarski2017high} (Z17). 

It is generally agreed that the radio-to-FIR spectrum of Cyg~X--1 is produced by its relativistic jets, and likely the GeV emission as well. Numerous studies dedicated to fitting high signal-to-noise X-ray spectra of Cyg~X--1 invoke the presence of a corona with hot, thermal electrons to upscatter soft disc photons up to $\sim100$ keV energies, as this is standard for most XRB hard-state models \citep{Tomsick_2013,duro2016revealing,Parker_2015,basak_2017,walter2017observations}. Furthermore, the companion of Cyg~X--1 is a high mass donor star, hence an additional black body (or even a more detailed stellar model) spectral component is required. 

The key differences between approaches centre primarily on the nature of the particle acceleration in the jets, the role of the jets at high energies, and the level of detail in the modelling of the jet properties. 

Constraining the contribution of the jets at high energies, and thus the total power requirements, hinges on the MeV polarisation and the $\g$-rays. Many of the prior works did not consider the MeV polarisation as a hard constraint. For those that did, R14 suggest that the synchrotron radiation from secondary electrons in the corona could explain the MeV tail. As we discussed above though, jet synchrotron is a more likely origin. Z14 explain the MeV flux as a result of jet synchrotron from primary electrons. They presented only a purely leptonic model and thus no TeV detection can be predicted. This choice thus places them in a regime with reasonable total jet powers. P15 manage to reproduce the MeV tail in a lepto-hadronic scenario with primary electron synchrotron radiation. This is similar to our lepto-hadronic model with $p=1.7$ but they use a much softer injected electron distribution. They manage to restrict the total proton power by making two assumptions discussed also above: first, protons are accelerated only from a minimum Lorentz factor of $\gamma_{p,{\rm min}}=100$ and second, the particle acceleration terminates at some distance from the jet base. None of these works though attempted to fit their free parameters to simultaneous data and perform statistical analysis, which may affect their conclusions.

\subsection{Perspective for CTA, HAWC, and LHAASO}
In Fig.~\ref{fig:spectrum with hadronic processes zoomed} we compare the results of the lepto-hadronic model with $p=1.7$ to the upper limits set by MAGIC after almost 100\,hours of observations \citep{ahnen2017search}. In addition, HAWC released its second catalog of TeV sources and a catalog of 9 Galactic sources after 1000\,days of operation, but Cyg~X--1 was not included in either of them (\citealt{HAWC2017firstcatalogue} and \citealt{HAWC2019tevcatalogue}, respectively). Thus, we also plot the sensitivity predicted by the HAWC collaboration for 5 years of operation \citep{HAWC2013sensitivity}. 

We plot the predicted sensitivity of CTA for TeV $\g$-rays, as well as the sensitivity of LHAASO \citep{bai2019large}, which mostly focuses on $\sim100\,$TeV. In the hadronic model with $p=1.7$, the TeV emission is dominated by the p$\g$ inelastic collisions between accelerated protons and synchrotron photons of the jet. The peak is at $20\,$TeV and the corresponding flux is expected to be $\rm{2\times10^{-12}\,erg\,cm^{-2}\,s^{-1}}$, significantly above the predicted CTA sensitivity for 50\,hours of observation from the north site. Moreover, the spectral index of this TeV emission is predicted to be positive and $\sim0.5$ for energies between 0.1--10 TeV (i.e. $F_{\nu}\propto \nu^{0.5}$).

An interesting aspect of our model is that the photomeson interactions dominate the pp collisions. The energy threshold of pp inelastic collisions, in general, is lower than p$\g$. Nevertheless, the number density of the target protons from the thermal wind of the companion star within the jet is constant up to $z\simeq a_{\star}$ regardless of the physics of the jets (see equation \ref{stellar wind}). On the other hand, the number density of the target photons of p$\g$ are highly model-dependent. For the hadronic models presented here the dominant target photons are the synchrotron photons of each jet segment. Consequently, in the case of the hard particle distribution ($p=1.7$) where the energy density of MeV photons is ten times higher than that of the soft particle distribution (see Fig. \ref{fig:spectrum with hadronic processes}), the p$\g$ process dominates the TeV band.

A detection of TeV photons and a measurement of the spectral index of this emission by forthcoming very high-energy facilities could therefore give further insights into the acceleration mechanism. Finally, regardless of the spectral shape, the detection of Cyg~X--1 from HAWC, and especially from CTA or LHAASO would exclude the possibility of purely leptonic jets for this source.

\section{Summary and Conclusions}\label{section: summary}
In this work, we present a new multi-zone jet model, based on the initial work of \cite{Markoff2005} and references above. We implement proton acceleration and inelastic hadronic collisions \citep[proton-proton and proton-photon,][respectively]{kelner2006energy,kelner2008energy}. We include the distributions of secondary electrons and $\g-$rays produced through pion decay. We further improve the existing leptonic processes with more sophisticated pair-production calculations \citep{coppi1990reaction,bottcher1997pair}, as well as take into account the proper geometry of the companion star as seen in the jet rest frame. With such enhancements, we can make more accurate predictions of the high energy phenomena related to astrophysical jets, particularly the non-thermal emitted radiation. 

Along with this new model, we present the first broadband, simultaneous data set obtained by the CHOCBOX campaign for Cyg~X--1 \citep{uttley2017}. This data set covers ten orders of magnitude in photon energy, from radio wavelengths to MeV X-rays. These bands are most susceptible to faster variability and hence simultaneous high-quality observations are beneficial to break model degeneracies. 

The keV-to-MeV spectrum of Cyg~X--1 exhibits significant evidence of linear polarisation. The keV spectrum shows low degree of linear polarisation \citep{Chauvin2018pogo,Chauvin2018pogo} but the 0.4--2\,MeV is highly polarised at a level of 50--70 per cent \citep{2011Sci...332..438L,jourdain2012separation,0004-637X-807-1-17,Cangemi2020longterm}. We interpret this high degree of linear polarisation in the MeV band as synchrotron radiation emitted by (primary) electrons accelerated inside the jets of Cyg~X--1 in the presence of a highly ordered magnetic field. Such non-turbulent, dynamically dominant magnetic fields are most likely associated with astrophysical jets. To achieve the required MeV synchrotron flux, we must inject a hard power-law of accelerated electrons with index of $p=1.7$.

We investigate the implications of the above assumptions for a purely leptonic and a lepto-hadronic scenario, performing statistical analyses to find the best fits to the CHOCBOX data set. Using an MCMC approach, we explore the parameter phase-space in order to constrain the parameters and minimize degeneracy. This paper is the first to compare a purely leptonic to a lepto-hadronic model for the case of XRB jets based on statistical analysis. 

We find that the jet geometry does not significantly differ between the two compared scenarios; the main differences are the TeV radiation and the power requirements. Only the hadronic model is capable of producing significant TeV emission detectable by the next generation $\g$-ray telescopes of HAWC, LHAASO and CTA. Interestingly, we find that the dominant hadronic process is the proton-photon interaction. This scenario however requires near-Eddington power in the accelerated protons, using the most basic assumptions.  We discuss ways around this issue but leave that for future work, in the case of a TeV detection.  Such a detection would be a game-changer for the field of XRBs, and support the possibility that Galactic CRs originate in more sources than only supernovae.

\section*{Acknowledgements}
We would like to thank the reviewer for the very helpful comments on improving the original manuscript. DK would like to thank Maria Petropoulou for fruitful discussions, and Ping Zhou and Thomas Russell for useful comments on the manuscript. DK, SM, ML, and AC were supported by the Netherlands Organisation for Scientific Research (NWO) VICI grant (no. 639.043.513). VG is supported through the Margarete von Wrangell fellowship by the ESF and the Ministry of Science, Research and the Arts Baden-W\"urttemberg. JCAM-J is the recipient of an Australian Research Council Future Fellowship (FT140101082), funded by the Australian government. This work is based on observations carried out under the project number W15BQ with the IRAM NOEMA Interferometer. IRAM is supported by INSU/CNRS (France), MPG (Germany) and IGN (Spain). This research made use of \verb+ASTROPY+ (\href{http://www.astropy.org}{http://www.astropy.org}) a community-developed core \verb+PYTHON+ package for Astronomy \citep{astropy:2013, astropy:2018}, \verb+MATPLOTLIB+ \citep{Hunter:2007}, \verb+NUMPY+ \citep{oliphant2006guide}, \verb+SCIPY+ \citep{2020SciPy-NMeth}, ISIS functions (ISISscripts) provided by ECAP/Remeis observatory and MIT (\href{http://www.sternwarte.unierlangen.de/isis/}{http://www.sternwarte.unierlangen.de/isis/}), and the CTA instrument response functions provided by the CTA Consortium and Observatory, see \href{http://www.cta-observatory.org/science/cta-performance/}{http://www.cta-observatory.org/science/cta-performance/} (version prod3b-v2) for more details.

%%%%%%%%%%%%%%%%%%%%%%%%%%%%%%%%%%%%%%%%%%%%%%%%%%\
\section*{Data availability}
The data underlying this article are available in Zenodo, at https://dx.doi.org/[doi]



\bibliographystyle{mnras}
\bibliography{CygX1.bib} % if your bibtex file is called example.bib

%%%%%%%%%%%%%%%%%%%%%%%%%%%%%%%%%%%%%%%%%%%%%%%%%%

%%%%%%%%%%%%%%%%% APPENDICES %%%%%%%%%%%%%%%%%%%%%
\appendix

%%%%%%%%%%%%%%%%%%%%%%%%%%%%%%%%%%%%%%%%%%%%%%%%%%

% Don't change these lines
\bsp	% typesetting comment
\label{lastpage}
\end{document}